\PassOptionsToPackage{hyphens}{url}
\documentclass[
]{article}
\usepackage{xcolor}
\usepackage{amsmath,amssymb}
\usepackage{iftex}
\ifPDFTeX
  \usepackage[T1]{fontenc}
  \usepackage[utf8]{inputenc}
  \usepackage{textcomp} 
\else 
\fi
\usepackage{lmodern}
\ifPDFTeX\else
\fi
\IfFileExists{upquote.sty}{\usepackage{upquote}}{}
\IfFileExists{microtype.sty}{
  \usepackage[]{microtype}
  \UseMicrotypeSet[protrusion]{basicmath} 
}{}

\usepackage{longtable,booktabs,array}
\usepackage{calc} 
\usepackage{etoolbox}
\makeatletter
\patchcmd\longtable{\par}{\if@noskipsec\mbox{}\fi\par}{}{}
\makeatother
\IfFileExists{footnotehyper.sty}{\usepackage{footnotehyper}}{\usepackage{footnote}}
\makesavenoteenv{longtable}
\usepackage{graphicx}
\makeatletter
\newsavebox\pandoc@box
\newcommand*\pandocbounded[1]{
  \sbox\pandoc@box{#1}
  \Gscale@div\@tempa{\textheight}{\dimexpr\ht\pandoc@box+\dp\pandoc@box\relax}
  \Gscale@div\@tempb{\linewidth}{\wd\pandoc@box}
  \ifdim\@tempb\p@<\@tempa\p@\let\@tempa\@tempb\fi
  \ifdim\@tempa\p@<\p@\scalebox{\@tempa}{\usebox\pandoc@box}
  \else\usebox{\pandoc@box}
  \fi
}
\def\fps@figure{htbp}
\makeatother
\usepackage{listings}
\usepackage{float}

\floatstyle{plain}
\newfloat{listing}{htbp}{lop}
\floatname{listing}{Listing}
\lstdefinelanguage{json}{
    basicstyle=\ttfamily\small,
    showstringspaces=false,
    breaklines=true,
    backgroundcolor=\color{gray!10},
    literate=
     *{0}{{{\color{blue}0}}}{1}
      {1}{{{\color{blue}1}}}{1}
      {2}{{{\color{blue}2}}}{1}
      {3}{{{\color{blue}3}}}{1}
      {4}{{{\color{blue}4}}}{1}
      {5}{{{\color{blue}5}}}{1}
      {6}{{{\color{blue}6}}}{1}
      {7}{{{\color{blue}7}}}{1}
      {8}{{{\color{blue}8}}}{1}
      {9}{{{\color{blue}9}}}{1}
      {:}{{{\color{purple}:}}}{1}
      {,}{{{\color{purple},}}}{1}
      {\{}{{{\color{orange}{\{}}}}{1}
      {\}}{{{\color{orange}{\}}}}}{1}
      {[}{{{\color{orange}{[}}}}{1}
      {]}{{{\color{orange}{]}}}}{1},
}
\usepackage[backend=biber,hyperref=true,sorting=none]{biblatex}

\DeclareFieldFormat{url}{\iffieldundef{doi}{\iffieldundef{eprint}{\url{#1}}{}}{}}

\usepackage[font=small,labelfont=bf]{caption}
\usepackage[export]{adjustbox}

\usepackage{arxiv}
\usepackage[T1]{fontenc}    
\usepackage{hyperref}       
\usepackage{booktabs}       
\usepackage{amsfonts}       
\usepackage{microtype}      
\usepackage{cleveref}       
\usepackage{mdframed,xcolor}

\definecolor{lightblue}{RGB}{173,216,230}
\newmdenv[
  backgroundcolor=gray!10,
  linecolor=gray!30,
  skipabove=\baselineskip,
  skipbelow=\baselineskip
]{codebox}
\usepackage{fancyvrb}
\fvset{fontsize=\scriptsize,formatcom=\ttfamily}
\usepackage{fvextra} 

\renewcommand{\arraystretch}{2.0}

\usepackage{titlesec}

\titleformat{\paragraph}[block]{\normalfont\normalsize\bfseries}{\theparagraph}{1em}{}
\titlespacing*{\paragraph}{0pt}{3.25ex plus 1ex minus .2ex}{1.5ex plus .2ex}

\usepackage{enumitem}
\setlist*{itemsep=0pt, parsep=0pt, topsep=0pt, partopsep=5pt}
\usepackage{bookmark}
\IfFileExists{xurl.sty}{\usepackage{xurl}}{} 
\hypersetup{
  pdftitle={A Multimodal GUI Architecture for Interfacing with LLM-Based Conversational Assistants},
  pdfauthor={Hans G.W. van Dam},
  hidelinks,
  pdfcreator={LaTeX via pandoc}}

\title{\protect\phantomsection\label{_Hlk204678888}{}A Multimodal GUI
Architecture for Interfacing with LLM-Based Conversational Assistants}
\author{Hans G.W. van Dam\\[-0.5em]uxx.ai}
\date{}

\begin{document}
\maketitle
\begin{abstract}
\protect\phantomsection\label{OLE_LINK4}{}Advances in large language
models (LLMs) and real-time speech recognition now make it possible to
issue any graphical user interface (GUI) action through natural language
and receive the corresponding system response directly through the GUI.
Most production applications were never designed with speech in mind.
\protect\phantomsection\label{OLE_LINK20}{}This article provides a
concrete architecture that enables GUIs to interface with LLM-based
speech-enabled assistants.

\protect\phantomsection\label{OLE_LINK11}{}The architecture makes an
application's navigation graph and semantics available through the Model
Context Protocol (MCP). The ViewModel, part of the MVVM
(Model-View-ViewModel) pattern, exposes the application's capabilities
to the assistant by supplying both tools applicable to a currently
visible view and application-global tools extracted from the GUI tree
router. \protect\phantomsection\label{_Hlk207008949}{}This architecture
facilitates full voice accessibility while ensuring reliable alignment
between spoken input and the visual interface, accompanied by consistent
feedback across modalities. It future-proofs apps for
\protect\phantomsection\label{_Hlk207008838}{}upcoming OS super
assistants that employ computer use agents (CUAs) and natively consume
MCP if an application provides it.

To address concerns about privacy and data security, the practical
effectiveness of locally deployable, open-weight LLMs for speech-enabled
multimodal UIs is evaluated.
\protect\phantomsection\label{OLE_LINK21}{}Findings suggest that recent
smaller open-weight models approach the performance of leading
proprietary models in overall accuracy and require enterprise-grade
hardware for fast
responsiveness.\protect\phantomsection\label{_Hlk205034539}{}

A demo implementation of the proposed architecture can be found at \textcolor{blue}{\url{https://github.com/hansvdam/langbar}}\end{abstract}

\section{Introduction}\label{introduction}

Integrating visual and linguistic interaction channels can substantially
increase the power, flexibility and accessibility of user interfaces.
Efficient multimodal interaction requires seamless integration between
graphical user interfaces (GUIs) and linguistic modalities. Ideally,
every action achievable through a graphical user interface (GUI) should
be equally attainable via voice commands. The GUI should then respond in
sync. Advances in large language models (LLMs) and thoughtful system
architecture now facilitate this goal \cite{wang2025guiagentsfoundationmodels, tang2025surveymllmbasedguiagents, wasti2024largelanguageuserinterfaces}.

This article is targeted at product managers, UX designers, and software
engineers who aim to develop high-quality, multimodal applications.
Beyond outlining high-level requirements and UX guidelines, this article
provides concrete architectural advice tailored explicitly for software
engineers, grounded in practical experience from real-world multimodal
systems.

Effective Human-Computer Interfaces (HCIs) should facilitate intuitive
interactions by seamlessly combining visual, linguistic, and gestural
modalities, much like the natural multimodal methods employed in
human-to-human communication
\cite{norman1988design, raman1997speech}.
Well-designed multimodal interfaces feel familiar and comfortable,
reflecting people\textquotesingle s ease in flexibly merging speech,
gestures, and visual cues according to context. Moreover, computerized
multimodal systems offer enhanced interaction possibilities due to their
dynamic graphical displays and real-time visual feedback.

This article discusses a software architecture that provides a natural
foundation for multimodal interaction. Key criteria for speech-driven
multimodal systems include

\begin{itemize}
\item
  Comprehensive coverage of application functionality via speech
  \cite{Reeves_2004}.
\item
  Accurate mapping of spoken requests into application actions.
\item
  High-quality real-time speech-to-text (STT) or direct audio
  interpretation via advanced LLMs
  \cite{ghosh2024gamalargeaudiolanguagemodel, ginart2024asynchronoustoolusagerealtime, openai_realtime_guide, liu2025voxtral, goel2025audioflamingo3advancing}.
\item
  Immediate and synchronized multimodal feedback (GUI plus TTS).
\item
  The ability to handle repair requests (self and other).
\item
  The capability to respond reliably to additional requests.
\end{itemize}

Achieving these goals in real-world implementations presents non-trivial
challenges in flexibility, maintainability, and user experience. Two
2023 articles inspired linking LLM function calling to mobile screen
functions, enabling elegant navigation via central app routing
\cite{vanDam2023SpeechInputLLMs, vanDam2023synergy}.
Experience in building such systems since then has broadened and refined
the architecture suitable for this kind of multimodality, which is
presented below.

This article begins by reviewing the core principles of GUI and
speech-based assistants. It then highlights recent advances in
speech-driven, OS-level assistants. Finally, it outlines a versatile
architecture that enables seamless connectivity between GUI-based
applications and embedded speech-driven assistants or external
assistants that operate at the OS level. Future OS-level assistants will
likely be dualistic, operating on direct visual observation of GUIs by
default and relying on application-provided semantics when these are
explicitly exposed, ideally through standardized mechanisms such as the
Model Context Protocol (MCP). Ultimately, this approach benefits both
developers and users: by exposing explicit semantics, developers can
enable assistants to offer more accurate and context-aware support,
while users experience smoother, more reliable linguistic interactions
across digital environments.

\section{Graphical User Interfaces}\label{graphical-user-interfaces}

GUIs have become pervasive because of their intuitive visual clarity and
structured presentation of tasks, enabling ease of use and quick
interactions for common scenarios. Understanding their strengths and
their limitations motivates the exploration of integrating
conversational speech capabilities, thus achieving a complementary and
enhanced user experience.

\subsection{Strengths of GUIs}\label{strengths-of-guis}

Key strengths of a GUI, compared to linguistic interaction, are the
following:

\begin{itemize}
\item
  Fast visual scanning: GUIs present a wide range of options and
  information simultaneously, enabling users, potentially, to quickly
  find what they need.
\item
  Direct and precise control: Actions such as clicking, dragging, and
  using sliders allow for highly accurate and intentional input.
\item
  Immediate visual feedback: Changes are reflected on-screen instantly,
  helping users understand the impact of their actions in real-time.
\item
  Easy discoverability: Menus, icons, and tooltips make available
  features and functions visible, reducing the need for guesswork about
  what the application has to offer.
\item
  Support for muscle memory: Frequent tasks become faster and more
  efficient as users build muscle memory for shortcuts and interface
  layouts.
\item
  Efficient multitasking: Multiple windows and tabs enable users to work
  on several tasks simultaneously without losing context.
\item
  Error prevention: GUIs can disable unavailable options and provide
  inline validation, minimizing the likelihood of user errors
  \cite{shneiderman2016designing}.
\end{itemize}

\subsection{Conversational Limitations of
GUIs}\label{conversational-limitations-of-guis}

Task-oriented spoken conversation consists of a mixture of open and
closed questions, assertions, confirmations, clarifications, and other
communicative acts, allowing flexible, context-sensitive interactions to
efficiently achieve user goals. For example, when ordering a pizza by
phone, the conversation might go like this:

\begin{quote}
\textbf{Staff:} ``Hello! How can I help you today?'' (open question)

\textbf{Customer:} ``I'd like to order a large vegetarian pizza.''
(answer)

\textbf{Staff:} ``Would you like extra cheese, mushrooms, or olives on
that?'' (closed question)

\textbf{Customer:} ``Extra cheese, please!'' (answer)

\textbf{Staff:} ``Are there any toppings you don't want?'' (open
question)

\textbf{Customer:} ``No onions, please.'' (answer)

\textbf{Staff:} ``So, that's one large vegetarian pizza with extra
cheese and no onions?'' (confirmation check)

\textbf{Customer:} ``Exactly!'' (confirmation)
\end{quote}

The design of GUIs often places system initiative at the forefront by
relying heavily on closed questions. While an open question, such as
``What would you like for dinner?'' allows for numerous possible
responses, the closed alternative, ``Do you want chicken, fish, or
vegetables?'' presents limited options, thus ensuring that the choices
offered can be readily fulfilled. However, the disadvantage of closed
questions is that users may become overwhelmed by the numerous options.
To mitigate this, modern GUIs segment interactions into narrower
contextual units, such as screens, panels, or menus, that directly align
with specific user intents. Designers face the recurring challenge of
balancing simplicity by restricting options versus ensuring completeness
by catering to less common use cases.

A familiar illustration of this trade-off can be seen in the GUIs of
home appliances, such as kitchen ovens. Although ovens today combine
diverse functionalities, such as microwaving, grilling, and conventional
oven heating, their user interfaces commonly fail to offer
straightforward access, making simple tasks like quickly baking
something excessively complicated. For example, a user may simply want
to preheat the oven to 180$^{\circ}$C, but is instead confronted with multiple
confusing buttons, icons, and menu options. This forces the user to
cycle through modes, adjust obscure settings, or consult the manual just
to accomplish what used to be a single, intuitive action.\\
The tightly focused nature of GUI interactions, evident in individual
screens or panels managing specific subtasks, supports cognition yet
simultaneously reduces flexibility. Ultimately, combining the precision
and focus of GUIs with the adaptability and openness of natural
conversation provides the richest solution, effectively empowering users
\cite{bieniek2024generativeaimultimodaluser}.

Examples of the challenge of balancing simplicity and flexibility can be
found in nearly every mobile app. For example, take a combined bank
transfer: ``Transfer 40 euros to each David and Sarah''. Most banking
apps do not offer such an option. In this case, the result can still be
achieved by making two transfers, requiring the user to repeat the
process for the second transaction. The absence of a combined transfer
option is due to concerns that adding such a feature would complicate
the transfer screen, where single transactions account for over 95\% of
daily use. The cost of incurring a cognitive burden on the default
scenario is slightly higher than the gain of providing a direct option
to make a double transaction.

GUIs also struggle with abstractions because they require users to
combine multiple filters and steps for complex tasks. For example,
finding ``all documents by Alice or Bob in the last month mentioning
both `budget' and `forecast'\,'' is hard to do in a GUI, but easy to
express in natural language as a single request.

\subsection{Linguistic Interaction with
GUIs}\label{linguistic-interaction-with-guis}

Blending GUI and linguistic interaction allows users to benefit from the
strengths of both modalities. While GUIs provide clarity, consistency,
and quick access to familiar functions, linguistic interaction enables
flexible, expressive, and intuitive communication. This synergy into a
multimodal UI makes it possible to combine the efficiency of direct
manipulation with the adaptability of natural language, resulting in
user experiences that are both powerful and accessible.

\section{Types of Multimodal UIs using
LLMs}\label{types-of-multimodal-uis-using-llms}

Multimodality encompasses a range of forms and dimensions, such as the
degree to which linguistics and graphics are interconnected and whether
the LLM interprets or generates graphics. In this article, four forms of
multimodal UIs are distinguished, as shown in Table
\ref{fig:typesOfMMUI}.

\begin{table*}[htbp]
\centering
\caption{Types of multimodal UIs distinguished by the kinds of information exchange between the GUI and the linguistic user interface}
\label{fig:typesOfMMUI}
\begin{tabular}{@{}c@{\hspace{1em}}c@{\hspace{2em}}p{\dimexpr\textwidth-2em-0.10\textwidth-3em\relax}@{}}
\midrule
1 & \includegraphics[width=0.10\textwidth,valign=t]{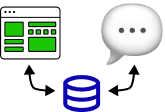} & \textbf{A GUI accompanied by a standalone chatbot interface on the side.} The chatbot bases its information and question answering on provided documentation by including in the prompt relevant text fragments that have been retrieved from documentation by a semantic search mechanism. This is called Retrieval Augmented Generation (RAG). The chatbot can also make API calls to perform actions. Although it can appear next to the GUI, it does not directly cooperate with it. \\
\midrule
2 & \includegraphics[width=0.10\textwidth,valign=t]{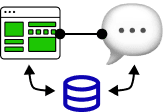} & \textbf{A GUI and a linguistic interface, with shared UI context.} The chat interface works in close cooperation with the GUI. Cohesion is hardcoded into the system. The semantics of the application is explicitly exposed, making it quite reliable. Moreover, the tight connection between linguistic and graphical user interface allows for high quality feedback in the form of graphics and speech, which is an essential component of conversational grounding. In addition, it can perform RAG and direct API calls just like type 1. \\
\midrule
3 & \includegraphics[width=0.10\textwidth,valign=t]{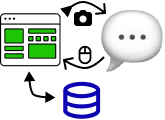} & \textbf{An agentic Multimodal LLM (MLLM)-based system based on observing the GUI.} The GUI is not directly integrated with the linguistic interface. Screen captures are fed to the LLM along with user utterances. The system can then perform actions in the GUI on the user's behalf by typing and clicking. These agents are sometimes referred to as Computer Use Agents (CUAs) and can execute tasks across multiple applications. In addition, such agents can perform RAG and direct API calls just like type 1. \\
\midrule
4 & \includegraphics[width=0.10\textwidth,valign=t]{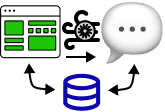} & \textbf{An agentic system that generates GUIs on the fly} as it sees fit. Users can interact with the generated GUI or issue speech acts to the system as they see fit. \\
\bottomrule
\end{tabular}
\end{table*}At the time of writing, most multimodal interfaces fall into the first
category: chatbot on the side. The second category is the main subject
of this article: strong cohesion between GUI and linguistic UI. There
have been some showcases for Type 4 (e.g.
\cite{youtube_v5tRc_5-8G4}), but they are still in an
early, rudimentary stage. Moreover, it is unclear whether such a
completely generic setup even makes sense beyond the inspirational value
of the concept.

Generic super assistants of type 3 are in active research and
development. They are designed to work on top of existing applications
without requiring modification to them. Moreover, these assistants can
execute actions that span multiple applications. The concept is very
powerful, but generic super assistants often struggle to reliably
complete tasks universally, among other factors, because it is
non-trivial to extract the semantics of the GUI solely from observing it
using screen captures.

In multimodal UIs of type 2, the semantic structure of the application
is provided to the LLM-based interface by the application itself rather
than extracted by an external GUI parser. The quality of interaction and
the reliability of task execution improve significantly when the
assistant has direct access to semantic structures explicitly defined by
the application provider. This article explains in detail how UIs of
type 2 (GUIs tightly coupled with a chat interface) can technically
expose their semantic structure, tailored to the current state of the
application. UIs of type 2 can operate standalone, with a dedicated
application-specific chat assistant, or expose their semantic structure
to an OS-wide super assistant of type 3.

Before delving deeper into multimodal interfaces of type 2 and exploring
precisely how they can seamlessly integrate with OS-wide super
assistants, it is important to first provide a good understanding of
generic super assistants of type 3.

\section{LLM-powered GUI Assistants}\label{llm-powered-gui-assistants}

\subsection{Computer Use Agents}\label{computer-use-agents}

\begin{figure}
\centering
\includegraphics[width=\linewidth,height=3.49048in,keepaspectratio]{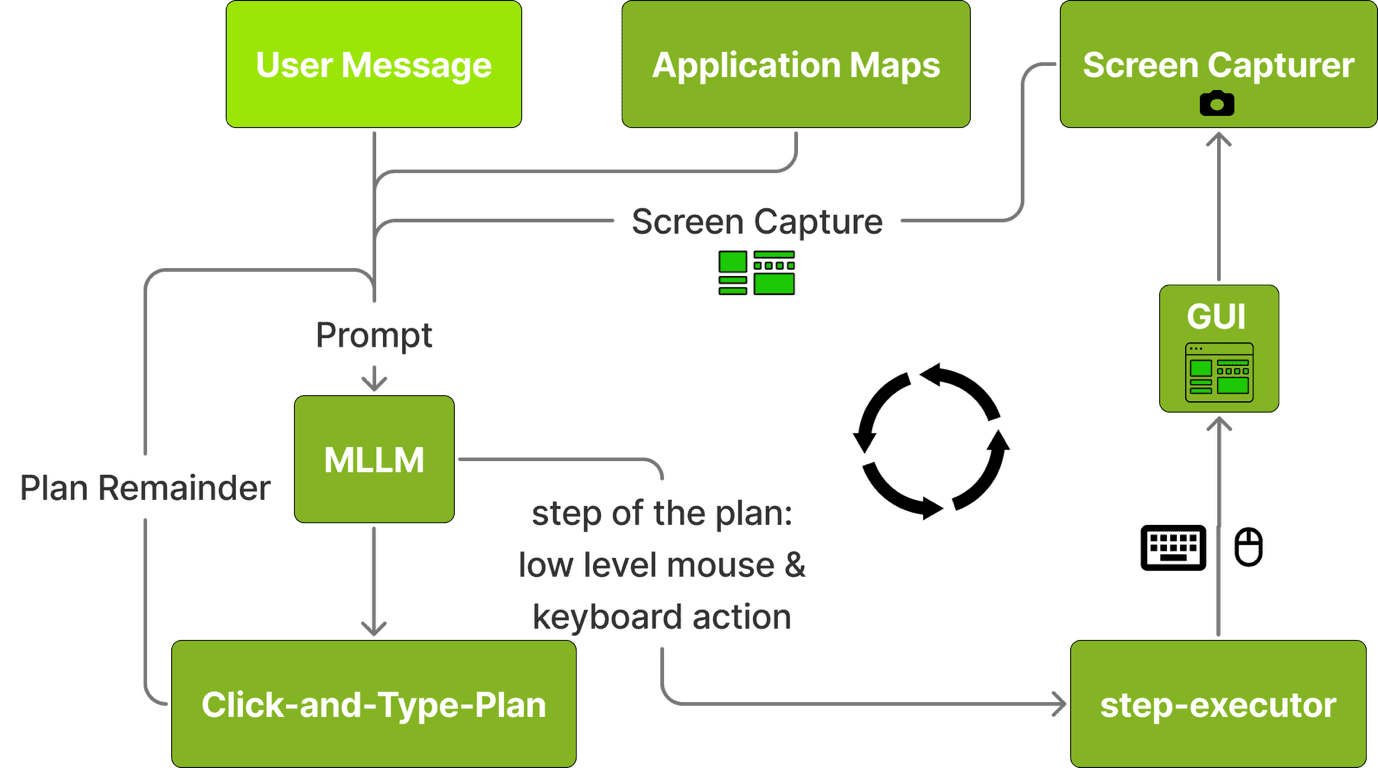}
\caption{Simplified flow of a CUA assistant.}
\label{fig:simplifiedCUAFlow}
\end{figure}

CUAs (type 3 in Table \ref{fig:typesOfMMUI}) are in active development
\cite{hu2024osagents, UI-TARS, autotab2025, chen2025osmapfarcomputerusingagents, yang2025macosworldmultilingualinteractivebenchmark, zhang2024lookscreensmultimodalchainofaction, zhang2025largelanguagemodelbrainedgui, tang2025surveymllmbasedguiagents, dsouza2024extended, nong2024mobileflowmultimodalllmmobile, li2025surveyguiagentsfoundation, buyl2024inherent, he2024webvoyagerbuildingendtoendweb, andreux2025surferhmeetsholo1costefficient, yan2025mcpworldunifiedbenchmarkingtestbed, zhang2025phigroundtechreportadvancing, wang2025opencuaopenfoundationscomputeruse}. Figure \ref{fig:simplifiedCUAFlow}
presents a highly simplified CUA workflow, which proceeds as follows.

When the user voices a request, the application creates a prompt that
includes a navigation map of the available applications and a screen
capture of the current view, together with agent instructions. The
Multimodal LLM (MLLM) creates an action plan consisting of button/menu
clicks, typing actions, or more semantic actions to be taken to achieve
the user\textquotesingle s goal. At this stage, the contents, the
precise layout, and dimensions of the current screen are known because a
captured screen has been sent to the LLM. The LLM outputs the first
low-level action to be taken in terms of hard x and y coordinates for a
click, potentially followed by some typing and a subsequent click in the
current screen. The remainder of the plan has to be refined into
low-level actions in subsequent cycles. After the step-executor clicks
or types on the GUI, the screen changes, a new screenshot is taken, and
the LLM determines subsequent low-level actions from the following
semantic action of the plan. This continues until the plan is fully
executed and the user's request is fulfilled.

In Figure \ref{fig:simplifiedCUAFlow}, for simplicity, planning is
portrayed as a one-shot prompt to an MLLM that combines the user message
with application maps and a screen capture to produce a plan. In
practice, planning approaches are typically much more fine-grained and
can handle tasks that span multiple applications or even the entire
operating system of a device \cite{vu2024gptvoicetaskeradvancingmultistepmobile, papoudakis2025appvlmlightweightvisionlanguage}. Rather than relying on a single-step
prompt, modern systems often decompose planning into multiple stages,
such as interpreting user intent and incrementally generating and
validating sub-goals. Aside from application maps, domain models, and
task templates may be used as input to the planning process
\cite{xiao2024flowbenchrevisitingbenchmarkingworkflowguided, niu2024screenagentvisionlanguagemodeldriven, zhang2024dynamicplanningllmbasedgraphical, kienle2025lodgejointhierarchicaltask, zhang2024lookscreensmultimodalchainofaction}. Moreover, as shown in Figure
\ref{fig:simplifiedCUAFlow}, after each step, the remaining plan is
re-evaluated using the screen capture resulting from that step, so the
plan can be adjusted if necessary.

Screen capturing can be performed using two primary methods. The first
method involves capturing a screenshot and then using an MLLM or visual
language model (VLM) to parse the image. The second method directly
scrapes and extracts information from a webpage's Document Object Model
(DOM) for processing by an LLM
\cite{browser_use2024}. Screenshots are typically
used for desktop and mobile applications. Still, even on mobile
applications, it is already possible to access the underlying GUI more
directly from outside the app using assistive technologies for
individuals with disabilities.

CUAs interpret the visual context of a GUI to determine the next
appropriate action. Ambiguities in the interface may constrain this
ability. For instance, consider a scenario where the agent encounters a
button labeled ``Cancel'' during a multi-step checkout process.
Depending on the context, clicking ``Cancel'' could either abort the
entire transaction or simply close a pop-up window without affecting the
ongoing process. Such ambiguity makes it challenging for the agent to
infer the correct action based solely on visual cues and textual labels
and may be unreliable when the agent lacks deeper contextual
understanding or explicit interaction data to resolve these ambiguities
\cite{xu2025webbenchllmcodebenchmark}.

To navigate to a different GUI context reliably, CUAs can use a map of
areas that are currently invisible. It can obtain a map of a GUI by
scraping an application, by clicking through its entirety once while
taking screenshots (see e.g. \cite{wu2025reachagentenhancingmobileagent}). Besides
generating maps, task templates can be crafted for an application by
hand or created semi-automatically, like in Google Project
Mariner\textquotesingle s `Teach and repeat'
\cite{project_mariner_2025} or Workflow Use
\cite{workflow_use2025}. In such systems, users
can generate semi-deterministic workflow templates that the system can
follow by demonstrating how to do it using a mouse and keyboard.
Compared to relying on purely automated, intelligent UI exploration,
user-driven demonstrations tend to produce workflows that are more
grounded and reliable. Other approaches to extracting workflows include
using software test code to generate workflows
\cite{weaver2024test2vareusingguitest}.

Mouse and keyboard actions executed by generic GUI automation-based
assistants naturally provide interactive feedback, enhance grounding in
user-agent communication, and facilitate user learning when users can
see the mouse pointer move and characters being typed. This is
particularly true when actions are performed at an easily comprehensible
pace. One might question the value of users learning actions that an
agent can already perform automatically. However, formulating tasks for
an agent still demands considerable cognitive effort, whereas performing
two familiar clicks can be performed almost subconsciously.

Despite the considerable promise of CUAs, there are many hurdles to
overcome \cite{yang2025mlatrustbenchmarkingtrustworthinessmultimodal}:

\begin{itemize}
\item
  \textbf{Reliability}: Ensuring robust execution across different
  interfaces remains a major challenge. Most commercial CUAs highlight
  their successful use cases but often fail to operate reliably in
  general. CUAs also struggle due to unpredictable GUI variations and
  dynamic element positioning, making consistent recognition and
  interaction difficult \cite{gao2024assistguitaskorienteddesktopgraphical, zhang2025phigroundtechreportadvancing}.
\item
  \textbf{Latency}: Capturing screens, parsing visual data through the
  LLM, and performing iterative action planning introduce delays. These
  delays can disrupt fluid user interactions, especially for complex,
  multi-step tasks.
\item
  \textbf{Safety}: Automated interactions may trigger unintended
  consequences, such as accidental data deletion, erroneous
  transactions, or security breaches, requiring rigorous constraints and
  careful validation before action execution
  \cite{maloyan2025investigatingvulnerabilityllmasajudgearchitectures, chen2025obviousinvisiblethreatllmpowered}.
\item
  \textbf{Privacy:} Cloud-based LLMs require sending potentially
  sensitive user data to external servers for processing, risking
  unauthorized access, leaks, and misuse. Additionally, these platforms
  may retain data, raising concerns about surveillance and compliance
  with privacy regulations \cite{luo2025dpolearningllmsjudgesignal}.
\item
  \textbf{Adaptability}: Generic assistants must cope efficiently with
  evolving software updates, changing layouts, or new GUI paradigms
  without extensive retraining or manual recalibration.
\item
  \textbf{Transparency}: Users need clear visibility into the reasoning
  behind the assistant's actions to build trust, particularly when
  automation involves sensitive or critical tasks.
\item
  \textbf{Scalability}: Navigating expansive application environments
  demands effective strategies for generating, managing, and updating
  GUI maps to prevent performance degradation and maintain seamless
  interactions.
\item
  \textbf{Action Mapping}: Accurately translating high-level user
  intents into precise sequences of low-level GUI actions requires
  sophisticated contextual reasoning, particularly for ambiguous
  requests or complex application logic.
\item
  \textbf{Contextual Interpretation}: Inferring accurate semantics
  purely from visual layouts is inherently non-trivial. Visual elements
  alone may lack contextual cues or meaningful labels, complicating the
  LLM's ability to reliably determine intended interactions.
\end{itemize}

\subsection{Hybrid assistance}\label{hybrid-assistance}

For linguistic interaction, the quality and reliability of system
responses are crucial to their acceptance by users
\cite{baughan2023mixedmethodsapproachunderstandinguser, lahoual2019users, goetsu2020differenttypesvoiceuser}. A reasonable quality of multimodal
interaction may be achieved for GUIs in general, as models and generic
techniques improve that primarily use screen captures and simulated user
input. For high-quality multimodal interaction, the linguistic assistant
must have a deep understanding of the application it assists. To achieve
this, the application can expose its semantics to the assistant through
an API \cite{lu2025axisefficienthumanagentcomputerinteraction, yang2024reactgeniedevelopmentframeworkcomplex, buyl2024inherent, zhang2025ufo2desktopagentos, zhang2025apiagentsvsgui, wasti2024largelanguageuserinterfaces}, ideally providing a modern connector
in a standardized format such as the Model Context Protocol (MCP).

\begin{figure}
\centering
\includegraphics[width=\linewidth,height=2.05951in,keepaspectratio]{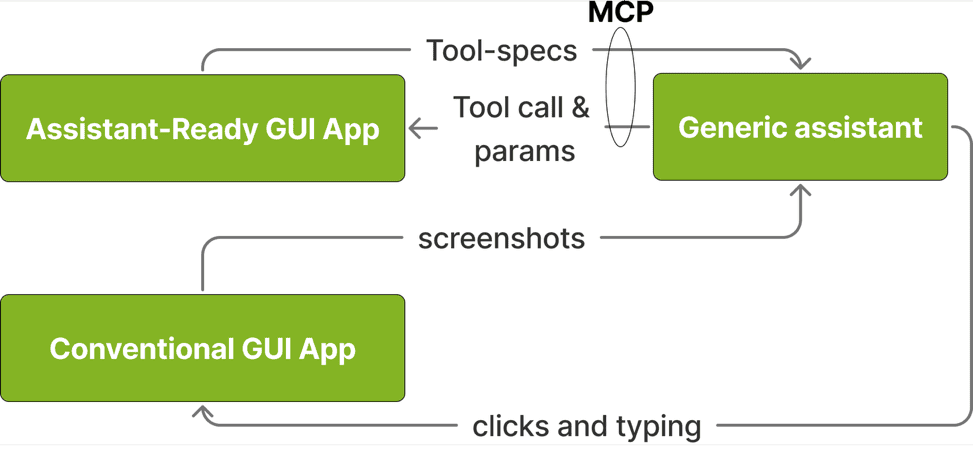}
\caption{Hybrid assistance, where enhanced GUIs explicitly expose their
semantics through callable tools and conventional GUIs are observed
through screenshots and operated on using mouse and keyboard
automation.}
\label{fig:hybridAssistance}
\end{figure}

The generic external screen-capture-based GUI interaction can serve as a
backup strategy for applications that lack a modern connector that
exposes their capabilities to the assistant. Hybrid assistance
connecting to both assistant-ready and conventional apps is shown in
Figure \ref{fig:hybridAssistance}.

This article focuses on hardcoding the cohesion between GUI and speech,
which offers application providers complete control to achieve
high-quality multimodal UX. It is the responsibility of the application
provider to build a user-friendly GUI. Although the GUI can, in
principle, be translated to a voice user interface (VUI) using generic
screenshot parsing, dedicated attention to this aspect of the user
interface leads to a more polished user experience. Moreover, textual
LLM calls, including tools, typically require fewer tokens than those
that include screenshots, have lower latency and cost, and higher
reliability \cite{yan2025mcpworldunifiedbenchmarkingtestbed}. Therefore, it is
advantageous for application developers to take full responsibility for
integrating the GUI and the VUI to work as one by providing an MCP API
to the generic assistant.

Hard-coded cohesion can be built using an architecture that aligns with
the hybrid approach mentioned above. This article will outline an
appropriate architecture to expose an application's capabilities and
handle responses from an LLM. This architecture enables developers to
create seamless integration between graphical and linguistic user
interfaces.

Besides describing an interface to generic OS level assistants, this
article also focuses on apps that deploy their own internal voice
assistant. In many use cases, this is desirable for increased control
and privacy. Moreover, generic assistance as described above is still in
an experimental stage.

When generic assistants mature, it becomes more important for
applications to provide a proper semantic connector to them. The
architecture discussed here facilitates that.

\section{Practical Use Cases of Speech-Enabled
GUIs}\label{practical-use-cases-of-speech-enabled-guis}

This section illustrates the potential benefits of integrating speech
assistance into GUIs. By presenting representative types of GUIs from
various domains, it highlights how speech-enabled interaction can
improve accessibility and usability. The following application types are
described with respect to their suitability for speech integration: a
mobile banking app for voice-activated navigation and data entry, a
shopping app for natural language search and task switching, a drawing
application for verbal manipulation of graphical objects, and a control
room dashboard for language-based command and assistance. These examples
are intended to illustrate possible use cases, providing context for the
subsequent section, which outlines a general architecture for
speech-enabling both mobile and desktop applications.

\subsection{Mobile Banking App}\label{mobile-banking-app}

\begin{figure}
\centering
\includegraphics[width=\linewidth,height=2.53004in,keepaspectratio]{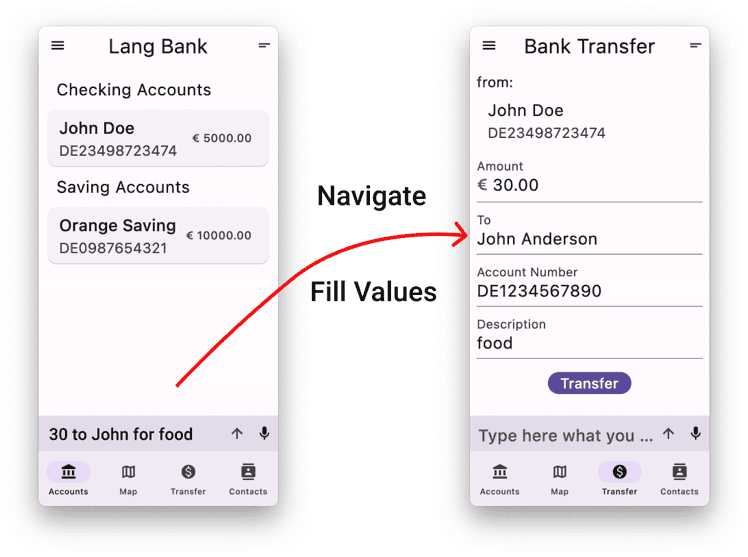}
\caption{A linguistic expression translated to app action: navigation to
the right screen and filling out parameters. Adapted from
\cite{vanDam2023synergy}}
\label{fig:navAndFill}
\end{figure}

Consider a banking app scenario as shown in Figure
\ref{fig:navAndFill}, where a user's spoken instruction is
automatically interpreted. This enables the interface to navigate
directly to the transfer screen and populate relevant fields with
extracted values such as the recipient and payment amount.

\subsection{Shopping App}\label{shopping-app}

When a customer uses a general assistant to shop in an online store, the
assistant can either show results directly or open the store's app and
perform the search there. It is often functionally preferable to have
the general assistant direct the search to the online store's App.
Allowing the store's app to manage the search can provide a more
optimized user experience in terms of presentation and checkout.
Integrating an LLM-powered search within the app enables users to refine
their queries through multiple interactions, combining natural language
and graphical interfaces. This approach separates understanding the
initial user intent from browsing products, which can improve usability.
If the user's query shifts away from product search, the system returns
control to the general assistant. This means users are not stuck within
the store's app. They can easily pivot to other tasks, like checking
their calendar or setting reminders, simply by changing the subject. As
a result, the experience combines the strengths of dedicated apps with
the flexibility of a general-purpose assistant.

\subsection{Drawing Application}\label{drawing-application}

\begin{figure}
\centering
\includegraphics[width=\linewidth,height=2.27325in,keepaspectratio]{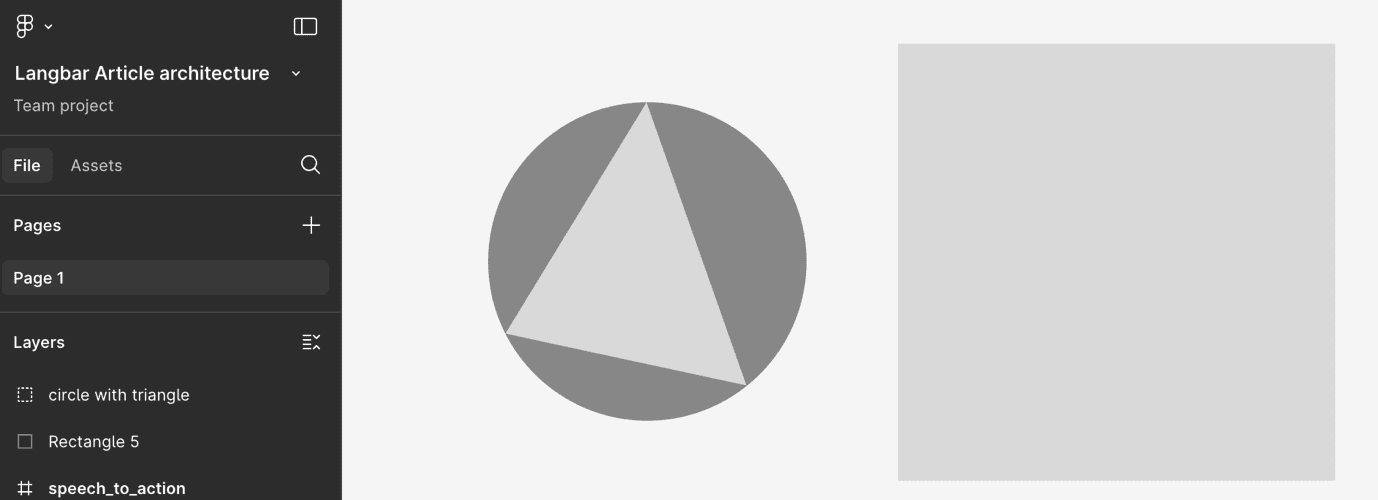}
\caption{The result of saying `Put the triangle to fit exactly into the
circle.' in a fictitious speech-enabled drawing application.}
\label{fig:triangleCircle}
\end{figure}

A Drawing application, as an example of a document-based application,
can greatly benefit from speech assistance, as it typically offers a
large number of tools, thereby putting a significant cognitive load on
the user. Consider the example in Figure \ref{fig:triangleCircle}.

It is challenging for a user to determine how to execute such actions in
the GUI. Being able to issue a simple command, as shown, would be
beneficial.

To date, several MCP servers have been created to interact directly with
drawings in a drawing application. For Figma design files, for instance,
``Cursor Talk to Figma''\\
\cite{lazuardi2025cursor} and
`Figma-Context-MCP' \cite{FigmaContextMCP} are
available. These MCP servers are still quite rudimentary, but more
importantly, they operate directly on the data structure of design
documents. It is promising to see how ``Cursor Talk to Figma'' allows
descriptive manipulation of graphical elements, even though the quality
is still relatively poor. Besides quality, for linguistic manipulation
to be useful, a closer integration with the GUI is necessary for a
better user experience. Real users need to be able to continue the
conversation in both the GUI and the design and switch seamlessly
between haptic and linguistic interactions. Drawing applications are
technically quite challenging to speech enable because of the broad
variety of functionality available within them. However, they are among
the most promising candidates for speech enablement from a user's
perspective, as they often have a steep learning curve.

\subsection{Control Room Dashboard}\label{control-room-dashboard}

\begin{figure}
\centering
\includegraphics[width=\linewidth,height=3.51268in,keepaspectratio]{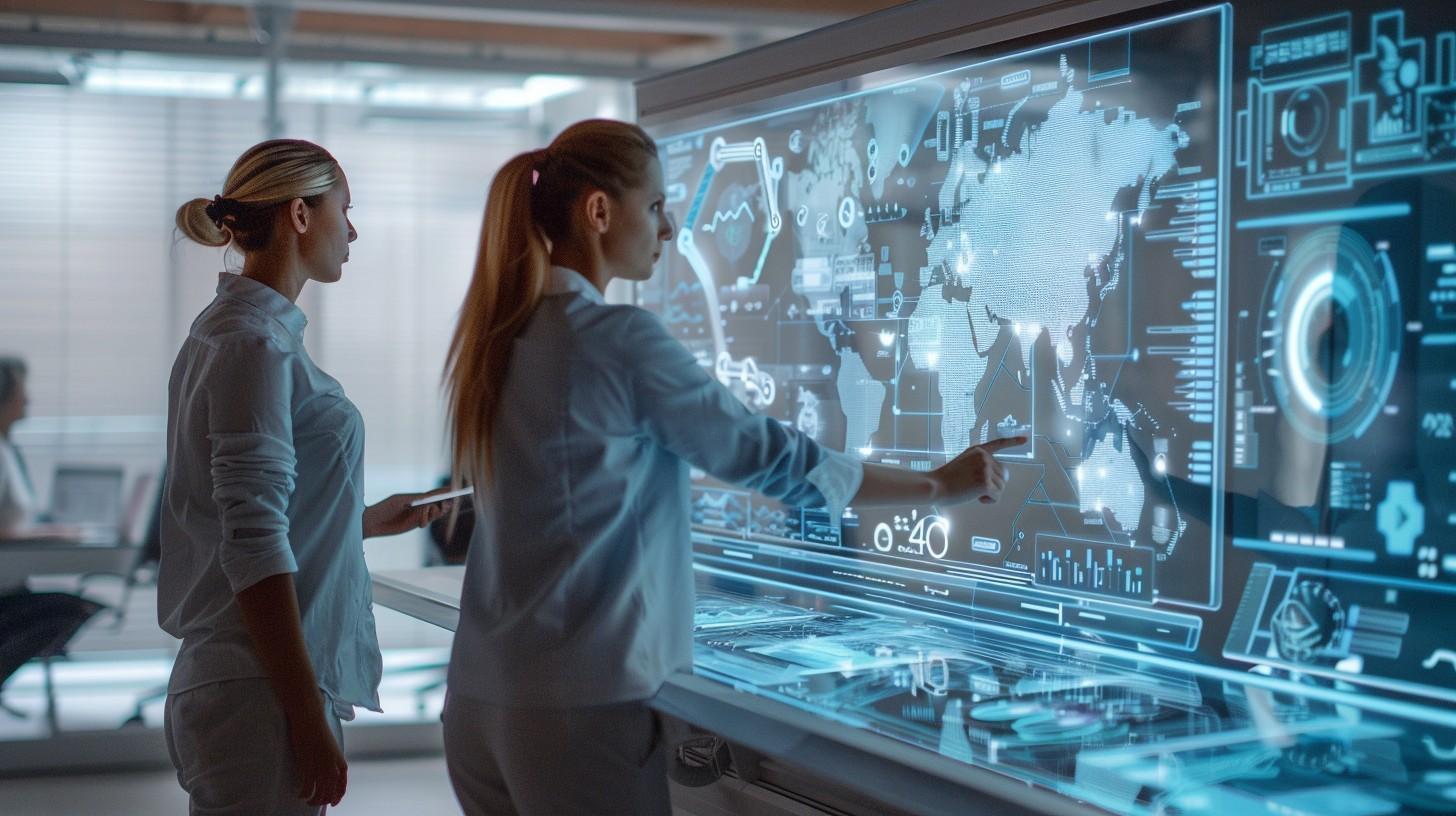}
\caption{A futuristic control room application, with many panels. It is
both complex and rigid, making speech enablement useful and
feasible.}
\label{fig:sampleDesktop}
\end{figure}

Control room dashboards are data-centric and constrained like a banking
app, but with significantly higher complexity. They are different from a
drawing application in that the elements display the state of systems
external to the application. Typically, only a limited portion of the
elements can be modified through the GUI. The users are generally
specialized and trained to know the system. Modifying the data is
typically more critical than changing the state of a document. Such GUIs
can greatly benefit from natural language interaction through LLMs in
two roles:

\begin{enumerate}
\def\labelenumi{\arabic{enumi}.}
\item
  As a primary tutoring assistant that can advise the user on actions to
  take.
\item
  As a power commander, it can issue commands over multiple entities at
  once. In this case, a two-stage commit for actions is important: after
  interpreting the user's request, the system should verbally state how
  it has understood the request and how exactly it intends to fulfill
  the request. Besides that, it should graphically highlight its
  intended action as a preview of the action itself. Actions are only
  executed after explicit confirmation from the user.
\end{enumerate}

\section{The Architecture of MCP-driven GUI
Applications}\label{the-architecture-of-mcp-driven-gui-applications}

\subsection{Introduction}\label{introduction-1}

A speech-driven system must have comprehensive knowledge and access to
all application functionalities to operate effectively. This article
introduces an architecture designed to fully support speech-enabled
interactions. Central to this architecture is the ViewModel, a component
responsible for orchestrating context, focus, and functionality
exposure, as detailed shortly. As a foundational step, it is crucial to
ensure universal speech-enabled coverage, allowing every part of the GUI
to be accessed efficiently via speech. Once this baseline is
established, designers can further optimize speech tools for areas of
the application where speech interaction provides use beyond the direct
mapping of the GUI. This approach ensures both comprehensive coverage
and optimal use of speech in contexts where it adds the most value.

The following section outlines the main concepts of a multimodal GUI
architecture. First, it explains the structure of navigation in response
to speech, then discusses how context influences the interpretation of
voiced requests. Next, it describes the architecture of multimodal apps,
followed by how this architecture supports two essential aspects of
communication: feedback and repair.

An implementation demonstrating the concepts discussed here can be found
at \url{https://github.com/hansvdam/langbar}.

\subsection{Key Constructs in Multimodal GUI
Architecture}\label{key-constructs-in-multimodal-gui-architecture}

\subsubsection{The GUI Tree Router}\label{the-gui-tree-router}

Modern GUI applications typically support centralized internal
navigation control. This is a great asset when enhancing them with
speech assistance, because it offers a central handle for an assistant
to cover the entire functionality of an application. To speech-enable a
GUI application, a developer can expose all its functionalities,
including detailed parameters, and ensure that all views and parameters
can be accessed through deep linking. Activations triggered by the LLM
or assistant are received centrally within the application and allow
direct access to any feature.

\begin{figure*}
\centering
\includegraphics[width=\linewidth,height=3.97996in,keepaspectratio]{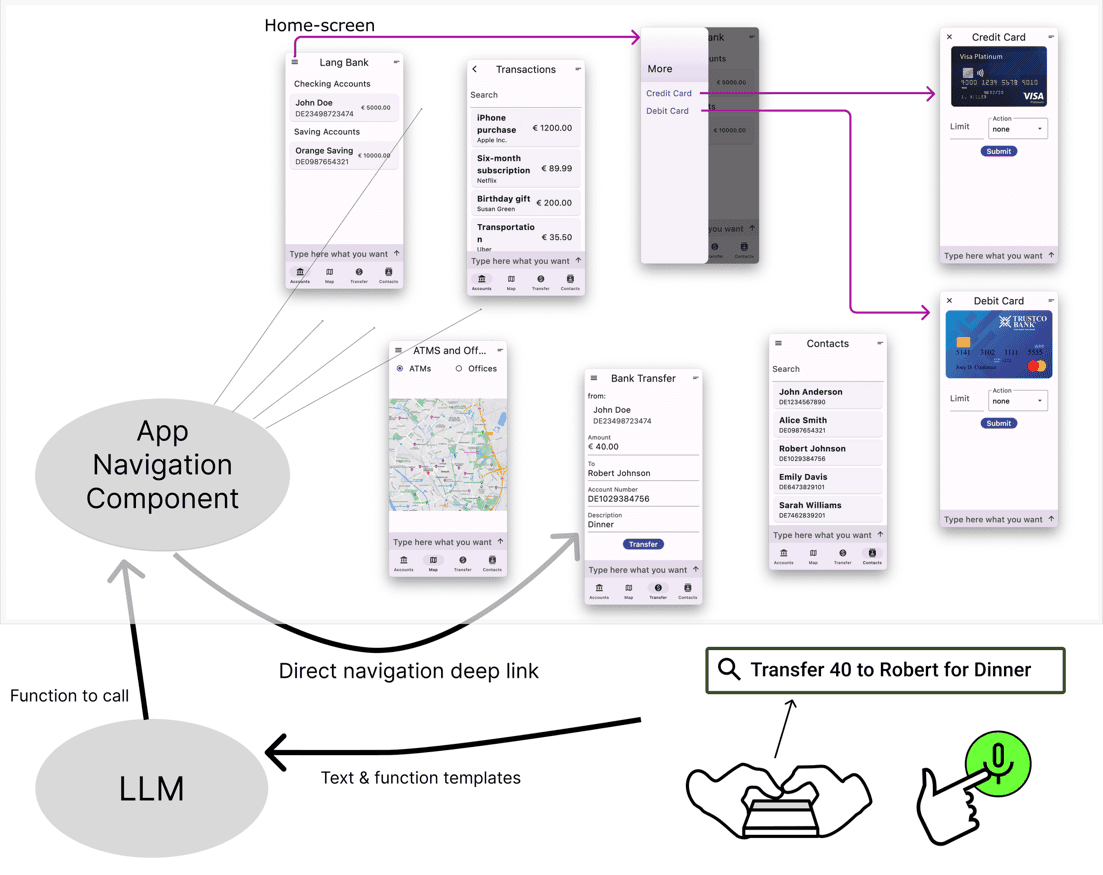}
\caption{Example of the navigation structure in a mobile app,
and how it leverages handling linguistic control. Adapted from
\cite{vanDam2023synergy}}
\label{fig:lingControl}
\end{figure*}

Central routers typically use hyperlink-like structures similar to REST
calls. Instead of making external API requests, these calls internally
activate specific application components and can populate parameter
fields in the GUI. This concept is illustrated for mobile apps in Figure
\ref{fig:lingControl}.

The user issues a request; the system constructs a prompt that consists
of the user\textquotesingle s request and a description of the app's
capabilities, in the form of structured documentation of all its main
views. In response, the LLM creates a tool call towards the assistant,
containing the view name and the parameters to fill in that view. Then,
the app or assistant translates this into a deep link that directs the
app to the correct view and fills in the parameters.

\begin{figure*}
\centering
\includegraphics[width=\linewidth,height=3.08333in,keepaspectratio]{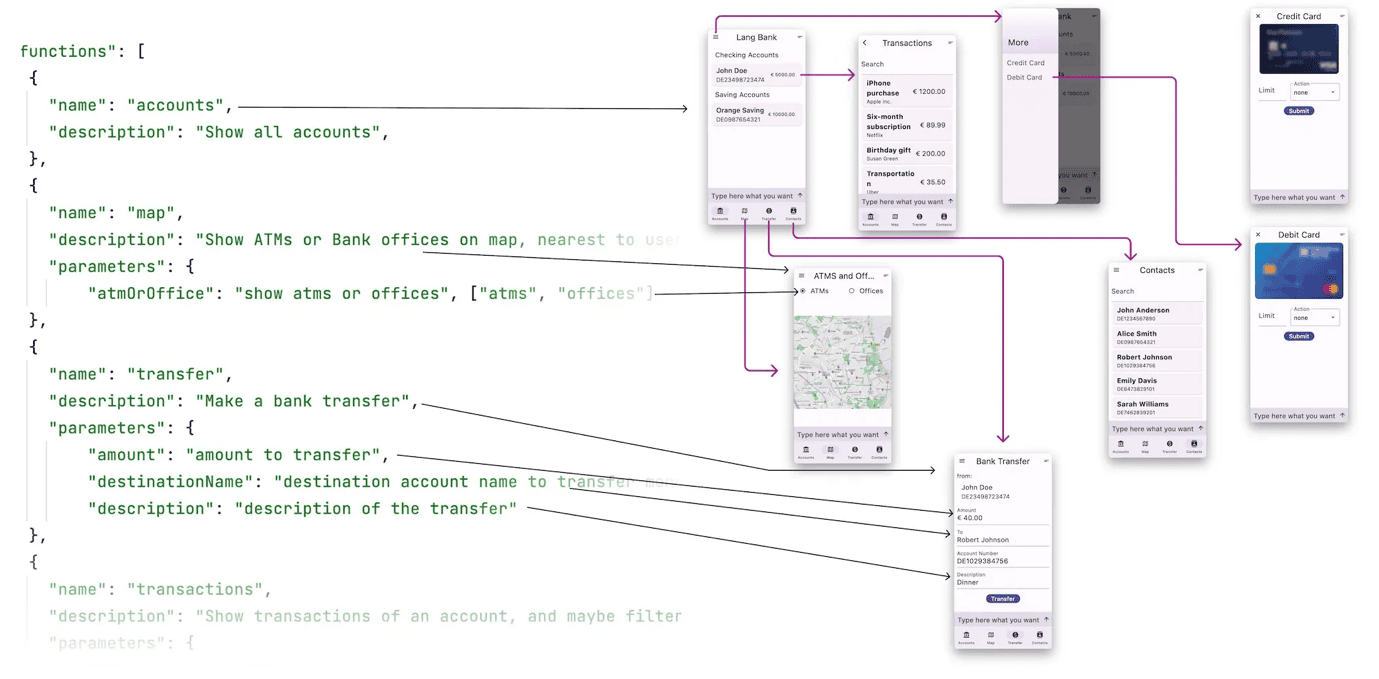}
\caption{Functions/tools for the LLM can map directly to
mobile screens. Adapted from
\cite{vanDam2023synergy}}
\label{fig:mobileDoc}
\end{figure*}

An example of the structured documentation of some screens in a mobile
app, in a simplified format, is shown in Figure \ref{fig:mobileDoc}.
This structure enables the app to understand global expressions,
navigate to the correct screen, and fill out the parameters.

Listing \ref{fig:documentedGoRouteItem} shows the Flutter code for the
part of the GUI tree router that navigates to the credit card screen and
sends the parameters to that screen
\cite{langbar}.\\
\begin{listing}[htbp]
\centering
\begin{lstlisting}[language=json, basicstyle=\scriptsize\ttfamily]
DocumentedGoRoute(
   name: 'creditcard',
   description: 'Show your credit card and maybe perform an action on it',
   parameters: [
     UIParameter(
       name: 'limit',
       description: 'New limit for the card',
       type: 'integer',
     ),
     UIParameter(
       name: 'action',
       description: 'Action to perform on the card',
       enumeration: ['replace', 'cancel'],
     ),
   ],
   pageBuilder: (context, state) {
     return MaterialPage(
         fullscreenDialog: true,
         child: LangBarWrapper(
             body: CreditCardScreen(
                 label: 'Credit Card',
                 action: ActionOnCard.fromString(
                     state.uri.queryParameters['action']),
                 limit:
                     int.tryParse(state.uri.queryParameters['limit'] ?? ''))));
   }),
\end{lstlisting}
\caption{Example Flutter code for a node in the GUI tree router that navigates to the credit card screen and sends the parameters to that screen \cite{langbar}}
\label{fig:documentedGoRouteItem}
\end{listing}The semantic documentation is embedded in the GUI tree structure,
ensuring a lasting one-to-one relationship and a single source of truth.

\subsubsection{Context}\label{context}

In human-computer conversation or interaction, the current context
directs the interpretation of the LLM. For instance, the sole user
expression ``raise the amount to 40'' has little meaning on its own, but
when the application shows the money-transfer screen of a banking app,
it is clear that the assistant should raise the amount to transfer. If
the user were on the credit card limit screen, the assistant should
raise the credit card limit. In mobile apps, the current screen forms
the context in which the expression is interpreted. Without that
context, a user expression would have to contain all the necessary
information to locate the right screen, even though that information is
already available because the right screen is already showing.

Context is added to the prompt and can contain elements such as:

\begin{itemize}
\item
  Conversation history
\item
  Current screen parameters
\item
  Application's capabilities (tool specification)
\item
  General System state
\item
  Expected output format
\item
  User data
\item
  Documentation
\end{itemize}

The exact buildup of the context and the creation of the final response
can span multiple calls to LLMs, allowing the LLM to reason from
high-level to detailed resolution.

\subsubsection{MVVM}\label{mvvm}

The architecture discussed in this article is based on the
Model-View-ViewModel (MVVM) pattern, in which the ViewModel plays a
central role. It bridges the gap between the user interface and the
underlying data. It acts as a mediator, ensuring a clean separation of
concerns and facilitating efficient communication. Most modern software
frameworks for applications with a GUI employ a declarative UI and
unidirectional data flow and have a concept quite similar to the
ViewModel, albeit under different names. This article does not propose
MVVM as superior to similar patterns with different nuances. It is used
as a representative architectural pattern for user-facing software,
where the gist of the multimodal flows described here is quite
applicable to other patterns as well.

The ViewModel prepares data in a display-ready format, a UI Model, which
differs from the raw source data (Entity). This separation ensures that
the UI does not directly manipulate the underlying data structure,
allowing for business logic tailored for UX, such as feedback, to
intervene.

GUIs are organized hierarchically, and the hierarchy of ViewModels that
form their backbone reflects the higher-level parts of this structure.
Mobile apps and Desktop apps typically have a slightly different
ViewModel hierarchy:

\par\medskip\noindent\begin{minipage}{\columnwidth}

\textbf{Mobile:}

ScreenViewModel

\begin{itemize}
\item
  ScreenPartViewModels
\end{itemize}

\end{minipage}

\par\medskip\noindent\begin{minipage}{\columnwidth}

\textbf{Desktop:}

WorkspaceViewModel

\begin{itemize}
\item
  WindowViewModel (MainWindowViewModel, etc.)

  \begin{itemize}
  \item
    RootViewModel (like a PageViewModel, DashboardViewModel, etc.)

    \begin{itemize}
    \item
      ChildViewModels (for controls, panels, tabs, etc.)
    \end{itemize}
  \end{itemize}
\end{itemize}

\end{minipage}

\subsubsection{The Role of ViewModels in Multimodal
Interaction}\label{the-role-of-viewmodels-in-multimodal-interaction}

In standard MVVM, the ViewModel forms the intermediary between the
user-facing View and the domain model. A domain model may be implemented
as a thin layer that exposes direct access to data, or as a thick layer
that incorporates multiple layers of business logic. With the
integration of a Voice User Interface (VUI), the ViewModel acquires an
additional role of combining the logic of the VUI and the GUI. In their
central role, ViewModels are also responsible for providing data for
both spoken and visual feedback after the user has voiced a request.

ViewModels determine which tools and commands are presented to the LLM.
They act as a publisher of and handler for the
application\textquotesingle s functionality. By organizing these tools
in order of relevance to the current context or screen, ViewModels
influence the LLM\textquotesingle s interpretation and response,
ensuring that the most appropriate actions are prioritized.

\begin{figure*}
\centering
\includegraphics[width=\linewidth,height=4.03589in,keepaspectratio]{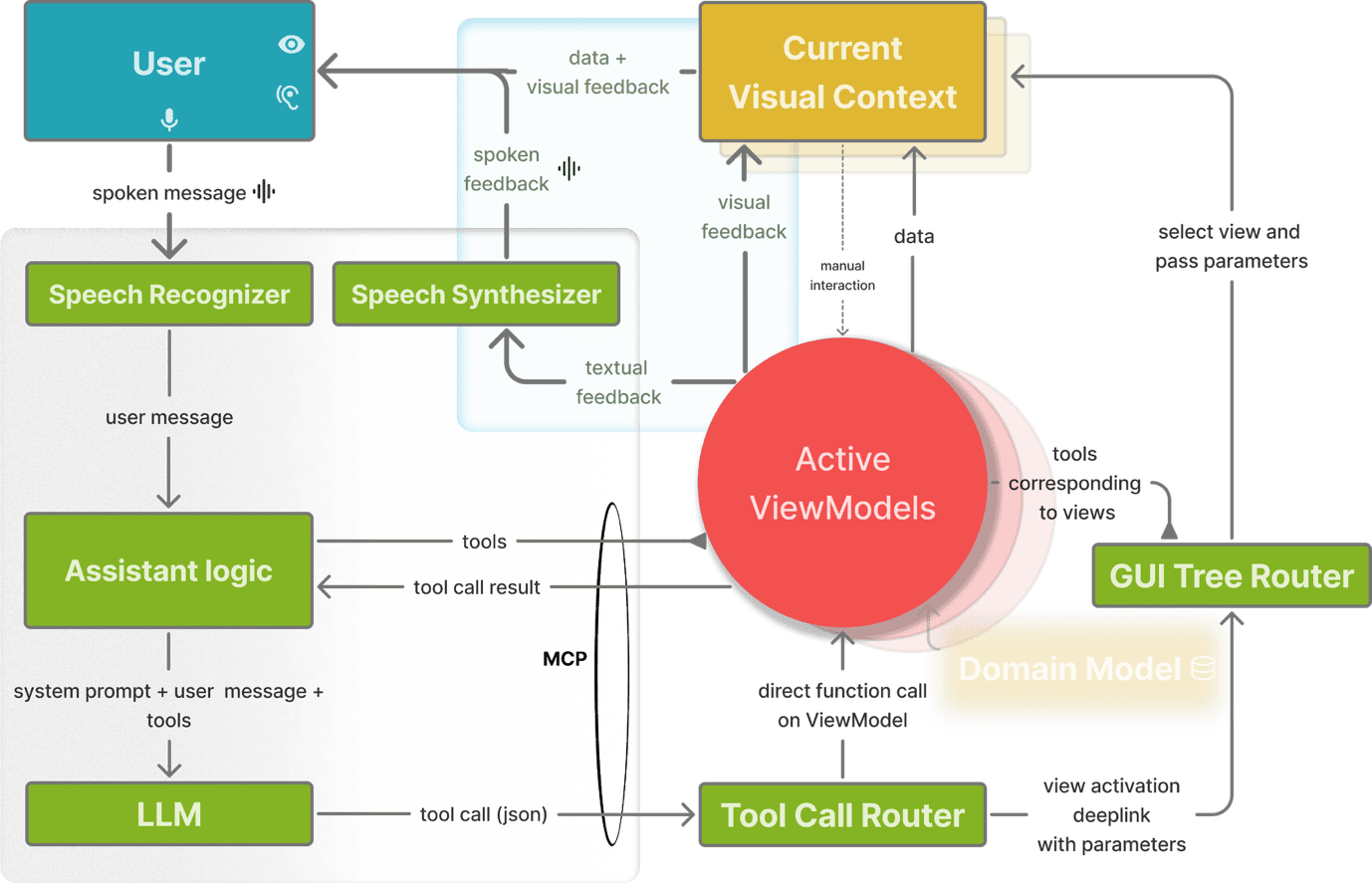}
\caption{The central role of the ViewModel in a multimodal
GUI.}
\label{fig:generalRoleViewModel}
\end{figure*}

The focus of a GUI is determined by what is currently visible on the
screen, which is referred to as the current visual context. In
multimodal interaction, ViewModels are not only the mediators between
the view and data. They also glue together the context for the LLM.
ViewModels can handle responses from an LLM directly when they apply to
the visual context they represent, rather than having the system route
them through the global application router. Figure
\ref{fig:generalRoleViewModel} shows the central role of the ViewModel
in multimodal applications that use MVVM.

In a mobile application the current visual context is formed by the
current screen. In a desktop application the current visual context is a
hierarchy formed by a workspace with windows and panels.

The tools exposed by the active ViewModels form a dynamic set that
changes depending on the context these ViewModels represent. This means
the app must notify the MCP client (the assistant) of these changes,
based on the context. It does this by issuing a
notifications/tools/list\_changed notification
\cite{mcp_tools_2025}.

Typically, a substantial part of exposed tools corresponds to navigation
through the GUI tree router. In addition, some tools correspond to a
direct function call on an active ViewModel and represent a set of
specific actions available for the current visual context. For these,
the ViewModel embeds the concrete handling function in the tool-object
so that it can be executed when that tool is called from the LLM. The
reference to the ViewModel is enclosed in the tool-object, i.e. it is a
closure. A tool in e.g. LangChain looks like this:

\noindent\begin{minipage}{\columnwidth}\begin{lstlisting}[language=json, basicstyle=\scriptsize\ttfamily]
class Tool(NamedTuple):
    """Interface for tools."""
    name: str
    func: Callable[[str], str]
    description: Optional[str] = None
\end{lstlisting}\end{minipage}
Where the \emph{func} entry represents the concrete handling function.

Tools created by the ViewModel pass a function to call as a native
function (Callable) object. When the LLM triggers a tool call, the
embedded function can be invoked directly. This works similarly for
tools that should trigger the GUI tree router. In the latter case, the
GUI tree router reference is enclosed in the function object.

\subsubsection{\texorpdfstring{Embedded Assistance vs. OS-Level
Assistance using MCP
}{Embedded Assistance vs. OS-Level Assistance using MCP }}\label{embedded-assistance-vs.-os-level-assistance-using-mcp}

When an assistant is created locally and is embedded into an
application, it the handling function is added to the tool, with either
a reference to the active ViewModel or the GUI tree router enclosed in
the function object.

\begin{figure*}
\centering
\includegraphics[width=\linewidth,height=3.58684in,keepaspectratio]{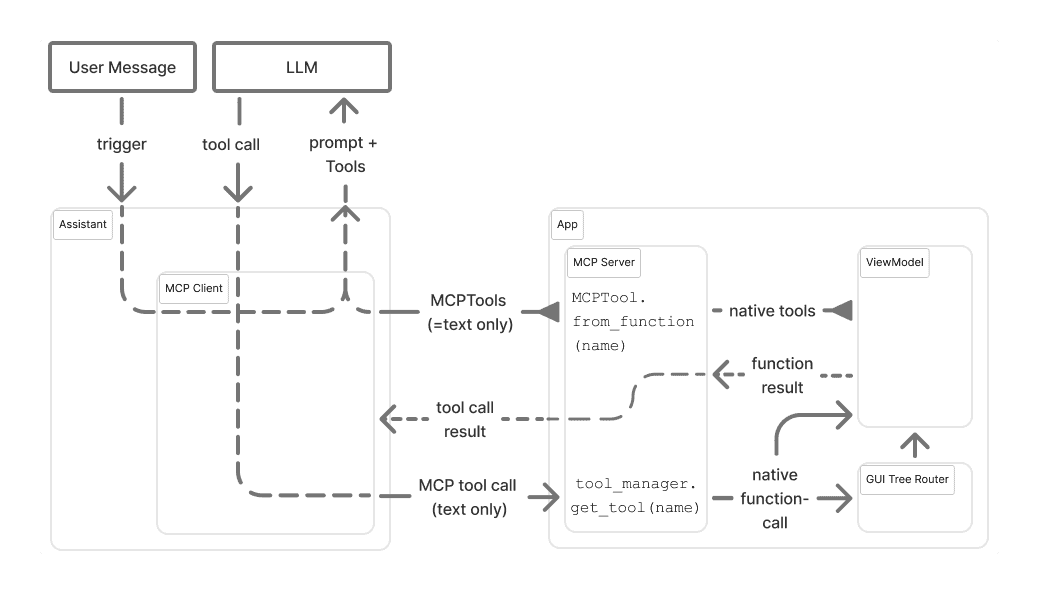}
\caption{Communication between application and assistant. An
application exposes an MCP server to a generic assistant, translating
native tools to MCP. The assistant's response is processed by the MCP
server and sent back into the ViewModel or GUI tree router.}
\label{fig:appAndMCP}
\end{figure*}

Through MCP, this works similarly, but tools and tool calls pass through
a client-server barrier as illustrated in Figure \ref{fig:appAndMCP}.
Tools to be called on the ViewModel or GUI tree router are passed to the
MCP server, including the function object that should be executed when
the corresponding tool is triggered. The MCP server maintains a list of
these tools in the current session. The function names and parameters in
this tool list are sent to the MCP client, i.e., the assistant. The
assistant calls the LLM, and when, in response, the LLM triggers a tool,
the function and parameter values are sent to the MCP server, i.e., the
app. The app performs a tool-name-lookup in the list that it maintains
and then calls the native function object that executes logic on the
ViewModel or GUI tree router.

A function call to the GUI tree router may lead to passing data to an
existing ViewModel or to a new ViewModel being activated. After
processing the data, the receiving ViewModel, new or existing, sends a
tool response back to the assistant, which represents a textual form of
what the user visually receives as a response to the GUI action
performed, such as the data displayed.

\subsubsection{Feedback}\label{feedback}

In human-to-human conversation, feedback in the form of paraphrasing
plays a crucial role in achieving \emph{common ground}
\cite{CLARK1989259, petukhova2015experimenting, glenn2022paraphrasing, books/others/91/ClarkB91, cf901d7e228d40ee82ec30e7b705d73e} or agreed-upon
truth.

When users issue a request to the system, they need to know how the
system has interpreted their request. This can be done both graphically
and verbally. Providing multimodal feedback gives the best
conversational grounding \cite{axelsson2022modeling}.
The need for the system to provide feedback is an important reason to
use the ViewModel as a central orchestrator. The ViewModel exposes the
tools and can also handle tool calls that operate directly on the
ViewModel. In some scenarios, it may be appealing to let the parameters
of speech tools act directly on the underlying data model, as the
ViewModel would inherently reflect the changes anyway. However,
bypassing the ViewModel as an input gate would make it impossible for
the ViewModel to determine the source of the changes and provide proper
feedback to the user. With the ViewModel in its central role as an
intermediary between the domain model and the graphical as well as the
voice UI, the ability to provide proper feedback for speech interaction
comes naturally.

\paragraph{TTS Feedback}\label{tts-feedback}

It is often beneficial if the system provides feedback in the form of
speech. This ensures that users clearly understand how their requests
have been interpreted and what actions the system is taking, especially
when the provided response does not exactly match the
user\textquotesingle s original intent. This feedback can verbalize the
tool call that the system will make. For Instance, when the user says:
``I want to speak to somebody at the counter'', the system may respond
saying: ``Showing offices on the map''. This is not what the user asked
for, but it is the best the app can offer in response to the request.

\paragraph{\texorpdfstring{Graphical Feedback
}{Graphical Feedback }}\label{graphical-feedback}

\begin{figure}
\centering
\includegraphics[width=\linewidth,height=4.10192in,keepaspectratio]{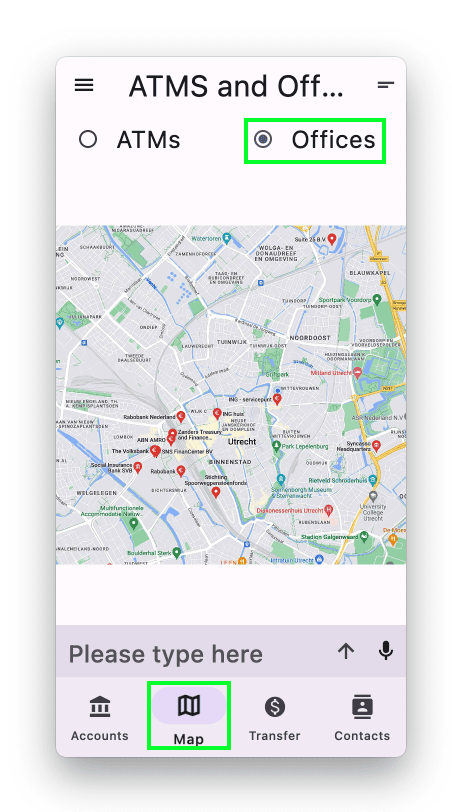}
\caption{An assistant graphically responds to a linguistic request, such
as ``Where can I visit a physical office?''. Explicit feedback is
essential for conversational grounding, emphasizing how the system has
responded to the request. The navigation button with the map and the
option button to show the offices are highlighted in green to ground the
interpretation of the request.}
\label{fig:highlighting}
\end{figure}

In response to an expression like ``Show offices on the map'', the
offices on the map are shown, just like when the user navigates there by
clicking in the GUI. Because the user did not actually click, the
assistance highlights the key graphical elements corresponding to its
interpretation, which the user would have tapped on if touch interaction
had been used. The navigation button with the map and the option button
to show the offices can be highlighted to ground the
user\textquotesingle s interpretation of their request. Such
highlighting can be accomplished by using a different color, an
enlargement, or an animation as shown in Figure \ref{fig:highlighting}.

The app does not have straightforward access to controlling the mouse
pointer, as it is managed by the operating system. This is an advantage
of the generic (desktop) GUI assistants described above. Their
automation is already based on mouse clicks, which means that the
actions of this form of assistance inadvertently provide feedback within
the conversation.

\subsubsection{Repair}\label{repair}

In human-to-human conversation, repair is crucial for achieving common
ground
\cite{CLARK19861}.
Speech, being more slippery than GUI communication, requires an
excellent repair mechanism. From a theoretical stance, repair in
task-oriented conversation can be divided in two types:

\begin{itemize}
\item
  Self-repair: Speakers correct their own speech.
\item
  Other-repair: Someone else corrects the speaker.
\end{itemize}

For a speech assistant it does not turn out to matter much which of the
two is at stake. A mistake in information transfer must be easily
repairable by speech. The speech assistant thus needs to be tuned so
that it reliably distinguishes repairs from continuations. If the user
utters ``Transfer 50 euros to Robert'', and subsequently utters ``No
40'', this should be flawlessly handled by the System, correcting the
amount to 40. Similarly, if the user utters: ``Transfer 40 to Robert'',
and the system understands ``Transfer 40 euros to Robin'', it should
handle a subsequent ``No to Robert'' and correct the previous mistake.
The fact that the first is a self-repair and the second is an
other-repair does not influence the required system behavior to correct
the error.

If both the assistant's message history and the tools are properly
configured, corrections like the one above are handled automatically.
This requires tool calls to include a parameter that clearly
distinguishes between updates and new instances, between corrections and
creations. This parameter has a default value of false and only needs to
be present in the LLM's response when the user indicates an intention to
create a new transfer, despite already working on a transfer on the
current transfer screen. For instance, when the user first utters:
``Transfer 40 to Robert'' and subsequently ``\emph{Also} transfer 50 to
Mary'', the LLM should respond to the second utterance with:

\noindent\begin{minipage}{\columnwidth}\begin{lstlisting}[language=json, basicstyle=\scriptsize\ttfamily]
{
  "name": "transfer",
  "properties": {
    "isNewTransfer": true,
    "destination": "Mary",
    "amount": 50
  }
}
\end{lstlisting}\end{minipage}
In this example, the isNewTransfer-parameter is added to the tool-call
structure to allow the app to make the distinction between corrections
and new instances.

\subsubsection{Handling Voice User Interface (VUI)
Parameters}\label{handling-voice-user-interface-vui-parameters}

The primary way to handle a user expression is for the LLM to translate
it into a tool call and have the central router deep link to the
applicable view and ViewModel, including parameters to be filled out.
This requires some modifications to most views compared to traditional
applications, where parameter values either come from underlying data or
changes to these values occur through user interaction with the GUI. In
the new speech-driven scenario, the call to open the screen also
includes parameters that must be entered into the GUI components on that
view.

When the user utters an expression that applies to parameters on a
currently visible view, there are two ways to implement the tool to
handle the expression. The simplest is to just let the central router
handle the tool call. The view is already there, and the default
behavior of most central routers is to just send the parameters through
to the already open view and ViewModel.

Another option is to have the ViewModel construct a special tool for
itself and include it in the list of tools it propagates to the
assistant. That way, it can, for instance, distinguish more accurately
between parameters that are already filled out on the screen and those
that are still empty. If they were filled from an interpretation of a
previous user\textquotesingle s expression or entered through GUI
actions, this information would be visible in the message history. But
if the data only came from data storage, a custom tool representing one
of the currently active views is also a good way to convey this data to
the LLM.

It is beyond the scope of this article to delve deeply into all
considerations affecting the choice between a local tool that calls
directly into a currently active ViewModel and the default approach of
letting the central router handle it.

\subsection{Mobile Apps}\label{mobile-apps}

This section tailors the general concepts discussed above more
specifically to mobile apps. Mobile apps typically have a very narrow
visual context, and only one screen ViewModel is active at any given
time. A screen here refers to a content area, and not to surrounding
navigation elements. For voice interaction, navigation elements are
implicitly handled via tools provided by the GUI tree router. For very
complex apps, the content area on a screen is divided into multiple
relatively separate sections. In such cases, the app has more
resemblance to a desktop app and should follow the hierarchical approach
described in the section about desktop applications below.

\subsubsection{The Role of ViewModels in Multimodal Mobile
Apps}\label{the-role-of-viewmodels-in-multimodal-mobile-apps}

\begin{figure*}
\centering
\includegraphics[width=\linewidth,height=3.63889in,keepaspectratio]{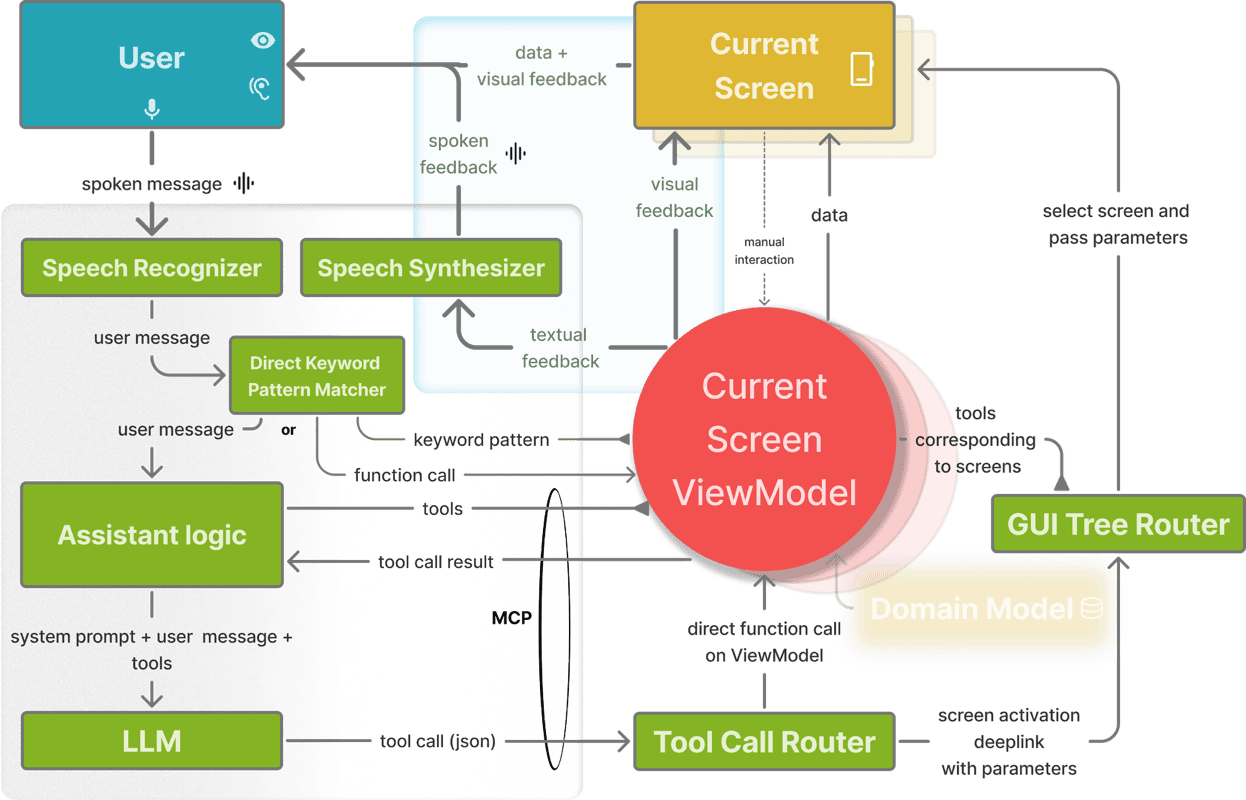}
\caption{In a multimodal mobile app, each screen has a
ViewModel that exposes the application semantics and provides graphical
and spoken feedback to the user in response to LLM tool responses.}
\label{fig:ViewModelMobile}
\end{figure*}

In a mobile app, the current visual context is formed by the currently
visible screen, and it is backed by the current ViewModel, as shown in
Figure \ref{fig:ViewModelMobile}.

The flow is as follows: the user voices a message, which gets translated
into text and passed to the LLM along with the tools received from the
current screen's ViewModel. The LLM responds with a tool-call JSON
object, which is sent either to the ViewModel or to the GUI tree router,
depending on the intended recipient.

The tools sent to the LLM are a combination of local functions for the
current view and global tools from the GUI tree router. Associated with
each tool in the application\textquotesingle s code is a corresponding
function object. When the LLM invokes a tool, the associated function
object is executed; each of these function objects inherently references
either the ViewModel or the GUI tree router, as these references are
encapsulated within their closures.

The ViewModel reads the global tools from the router and orders them
strategically, typically placing the function corresponding to the
current screen at the top. LLMs are more likely to select the first tool
in their prompt due to ``positional bias''
\cite{shi2025judgingjudgessystematicstudy}. In an app, it is most likely that a
user expression refers to something that applies to the currently
visible screen.

Tool call results are sent back from the LLM to the assistant to be
added to the conversation history and can serve as information transfer
units for multi-step tasks. The result of a tool call for the user is
typically a GUI displaying something on the screen. The tool call result
sent back to the assistant logic by the ViewModel is a textual
representation of what the user sees on the screen.

If the speech assistant is embedded in the application, communication
between the ViewModel and the assistant occurs directly within the
application\textquotesingle s code. If an external generic assistant
handles voice interaction, the application communicates through MCP.

\subsubsection{Direct Keyword Matching}\label{direct-keyword-matching}

The multimodal architecture discussed in this chapter harnesses
applications with the sophisticated language interpretation skills of
LLMs. Additionally, an application may want to expose short commands
that follow a simple pattern. These commands do not require the LLM for
interpretation and can be triggered immediately after the speech
recognizer transcribes the audio. This enables the assistant to respond
more quickly to some commands, although it only works if the user is
familiar with their existence. Such a command pattern can be expressed
using a regular expression. An example is the regular expression
``(\textbackslash b\textbackslash w+\textbackslash s+)?{[}bB{]}ack''
that is triggered by expressions like ``go back'', ``click back'',
``navigate back'', or simply ``back'' or ``Back''.

Using keyword pattern-matching this way bypasses the LLM for short
navigation commands. It is an optimization that could also have been
tackled using a plain Tool.

When a user expression is intercepted using a direct keyword matching
pattern after the speech recognizer, the LLM is bypassed completely.
Bypassing the LLM is not part of MCP. However, it is possible to
integrate keyword matching into the MCP protocol as a tool and have the
MCP client, i.e. the assistant, remove that tool from the tool list to
match against the speech recognizer\textquotesingle s result directly.
If a match is found, the response is sent back directly to the
application. If no match is found, the remaining MCP tools are sent to
the LLM.

If the assistant does not treat the regular-expression-matcher tool as a
special case, the tool can also be used straight away by the LLM to
produce the same result, albeit via a longer and slower route. The
function structure for encoding regular expressions as an MCP tool is
shown in Listing \ref{fig:regexTool}.\\
\begin{listing}[htbp]
\centering
\begin{lstlisting}[language=json, basicstyle=\scriptsize\ttfamily]
{
  "name": "match_user_input",
  "description": "Check user input against specified regular expressions and return matched input.",
  "parameters": {
    "type": "object",
    "properties": {
      "user_input": {
        "type": "string",
      },
      "regexps": {
        "type": "array",
        "description": "regular expressions for matching.",
        "items": {
          "type": "string",
          "enum": [
            "(\\b\\w+\\s+)?[bB]ack",
            "(\\b\\w+\\s+)?[fF]orward"
          ]
        }
      }
    },
    "required": ["user_input", "regexps"]
  }
}
\end{lstlisting}
\caption{Tool that can be used as a regular expression matcher for forward and backward commands in an LLM. This is only a fallback strategy for when the assistant does not understand that it should bypass the LLM in case of a match.}
\label{fig:regexTool}
\end{listing}Using MCP to communicate literal voice commands to the assistant is a
hack. It is efficient because it removes the need for dedicated
assistance channels, but it depends on some degree of name agreement and
extends MCP usage beyond its original intent.

\subsubsection{The Challenge of Handling Historical
Context}\label{the-challenge-of-handling-historical-context}

Ideally, the history of the entire conversation between the user and the
system is kept as context for interpreting the next user expression.
LLMs, however, can easily get confused by historical messages,
especially when the subject changes during the conversation. Quite
often, after such subject changes, the history of the conversation
before that is irrelevant. However, retaining that history can confuse a
language model when interpreting the user's current input. Especially
for weaker LLMs, it is beneficial to clear the history before any new
screen is opened to avoid confusing the model.

When manual GUI interactions trigger a screen transition, the new screen
identifier is appended to the LLM history to ensure accurate
interpretation of subsequent user input. Integration with external
speech assistants via MCP could achieve this by leveraging
MCP\textquotesingle s sampling mechanism through message creation
(server.createMessage). However, the MCP specification mandates explicit
user consent to inject user message updates into the conversation
history \cite{mcp2025}.
Therefore, an alternative approach is required.

The following two MCP-compliant solutions can be used to expose relevant
manual GUI interactions for inclusion into an OS-wide assistant's
message history:

\begin{enumerate}
\def\labelenumi{\arabic{enumi}.}
\item
  \textbf{Use of Silent Context Resources}: Rather than injecting events
  directly into the conversation, GUI state changes can be exposed
  programmatically via read-only MCP resources (e.g., /last-gui-events).
  This method allows assistants to retrieve contextual data as needed
  through existing MCP-approved tool calls, eliminating unnecessary chat
  clutter and avoiding repetitive consent prompts.
\item
  \textbf{Assistant Role Attribution}: Alternatively, GUI-triggered
  messages can be introduced with "role":"assistant". MCP clients
  generally auto-approve assistant-generated drafts, acknowledging them
  as non-user-authored. Careful, non-actionable wording is critical here
  to mitigate any injection risks.
\end{enumerate}

MCP is not intended for the server side, i.e., the application, to take
the initiative in the conversation. Future versions of assistant
platforms may adopt new mechanisms, designed explicitly to streamline
and safely integrate GUI interactions into assistant conversation
histories.

\subsubsection{Rare Use Cases}\label{rare-use-cases}

GUI screens for rare use cases that would clutter the navigation
structure may still be relevant enough to include in speech interfaces.
For example, in most banking apps, a screen for combining payments to
multiple beneficiaries is lacking, even though it would be convenient
for users who need it. Presenting too many options in a main navigation
menu can lead to clutter and an increased cognitive load. Moreover, rare
complex options are generally difficult to explain. It is better for
such a rare case to have the user execute a money transfer twice than
clog up the navigation menu with a combined transfer option.

In a Voice User Interface (VUI), options are not presented through a
navigation menu because the initiative lies with the user
\cite{wasti2024largelanguageuserinterfaces}. The absence of navigation allows for
catering to less frequent use cases with specialized GUI screens,
without detrimental effects on the usability of the GUI as a whole
\cite{lu2025axisefficienthumanagentcomputerinteraction}. This flexibility enables designers
to address complex or niche tasks directly, expanding the interface's
functional reach while keeping the core experience streamlined and
user-friendly.

\subsubsection{Langbar: GUI assistant and Chatbot in
One}\label{langbar-gui-assistant-and-chatbot-in-one}

\begin{figure}
\centering
\includegraphics[width=\linewidth,height=2.9626in,keepaspectratio]{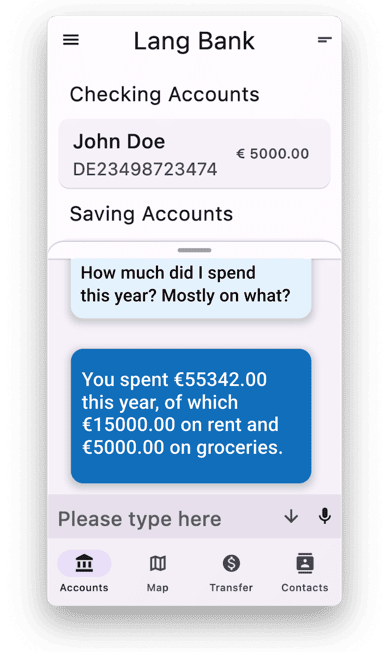}
\caption{Langbar\textquotesingle s expandable history panel
automatically opens for pure linguistic interaction without wasting
screen real estate when the system responds graphically to linguistic
requests.}
\label{fig:historyBar}
\end{figure}

When an app has its own local LLM assistant, it must reside somewhere on
the screen without occupying excessive screen real estate. A solution
for this is introduced in
\cite{vanDam2023synergy}
in the form of the ``LangBar'' concept, which combines a linguistic
interaction field that allows for entry by keyboard as well as speech,
with an expandable history panel as seen in chatbots, as shown in Figure
\ref{fig:historyBar}.

Besides all the functionality that is available through GUIs, it is
possible to offer a wider set of functionality to the user, e.g., via a
chatbot, like in the abovementioned LangBar. Moreover, user data, such
as address books or transaction history, can be transferred to the agent
using queryable MCP tools (e.g., using
\cite{genai_toolbox} or
\cite{aaronsb_google_workspace_mcp}
(get\_workspace\_contacts)). This enables assistants to provide more
personalized and context-aware support by using direct access to
relevant user data and services beyond what is typically accessible
through the GUI alone.

\paragraph{GUI History Trace}\label{gui-history-trace}

\begin{figure}
\centering
\includegraphics[width=\linewidth,height=1.69685in,keepaspectratio]{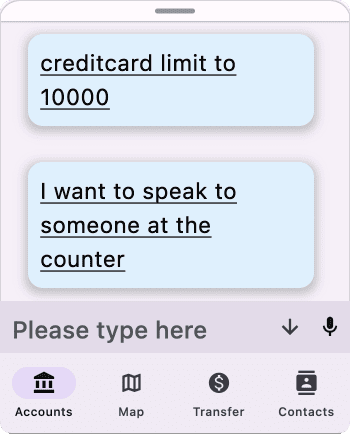}
\caption{The history panel can expand to display past actions as
clickable links, allowing users to repeat those actions.}
\label{fig:repeatAction}
\end{figure}

Requests that the assistant responded to in the form of a GUI action can
be added to the visible history in the form of a hyperlink that executes
the corresponding deep link through the central router of the app, as
shown in Figure \ref{fig:repeatAction}.

The history panel is only shown when the user explicitly expands it or
when the assistant answers using text or speech.

\subsection{Large Desktop
Applications}\label{large-desktop-applications}

\begin{figure*}
\centering
\includegraphics[width=\linewidth,height=3.55376in,keepaspectratio]{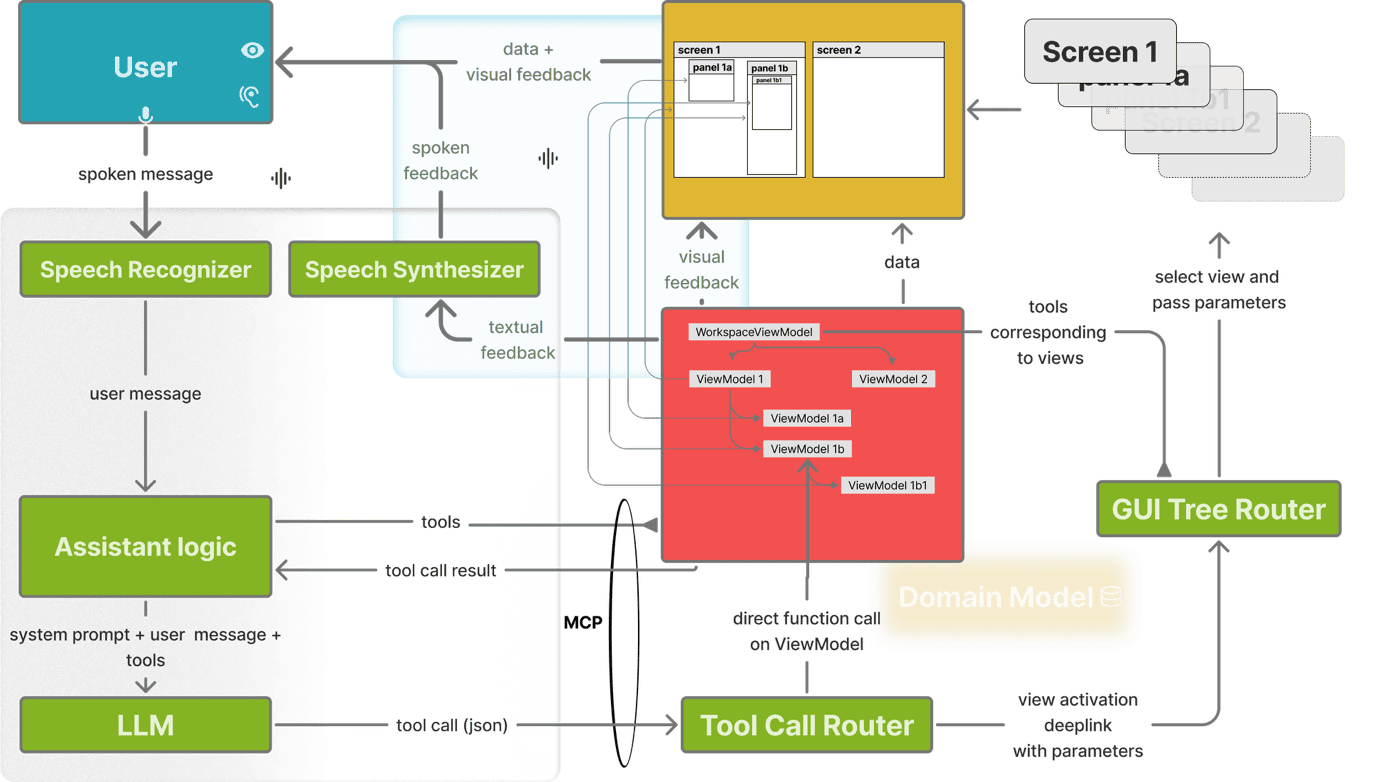}
\caption{ViewModels in desktop apps are hierarchically
ordered, reflecting the GUI structure. ViewModels for child views
propagate their tools upwards. The workspace ViewModel is ultimately
responsible for exposing all tools. Child ViewModels can handle the tool
responses of their own tools.}
\label{fig:viewModelDesktop}
\end{figure*}

The preceding example of a banking app demonstrated the simplest case of
the proposed GUI architecture. This section expands this method to
larger-scale desktop applications, such as control room user interfaces.
In desktop GUIs, multiple workspaces, screens, windows, panels, and
interactive elements coexist simultaneously. Unlike a banking
application, which typically possesses a single, clearly defined current
view, complex applications have overlapping contexts and multiple
interaction points. Focus and priority, rather than a singular current
screen, determine the user\textquotesingle s visual and operational
context. For example, panels may be highlighted through user selection
or cursor positioning, establishing a hierarchy of active contexts. The
general architecture for a multimodal desktop application is shown in
Figure \ref{fig:viewModelDesktop}.

The WorkspaceViewModel has access to the ViewModels for the screens or
windows, which in turn have access to the ViewModels of the panels. The
WorkspaceViewModel composes a set of tools for the assistant by
combining global (navigation) tools with the tools it gets from the
ViewModels of the windows. They, in turn, compose their toolset from
tools that operate at the window level and tools they obtain from
ViewModels at the panel level. Every ViewModel can have its own specific
take on what it exposes to the upper level, so the tool-exposure
architecture is fully customizable.

A challenge for large desktop applications in general is that they may
offer a toolset with too many choices for the LLM. Like humans, LLMs are
notoriously bad at making choices from a large set. If one continues to
add complex tools to the toolset, it is essential to maintain thorough
evaluations (tests). As the toolset grows and more obscure tools are
added, the most important tools should remain easily accessible and
should still be chosen when appropriate. When accuracy deteriorates, one
needs to prioritize the toolset. It is good practice to prioritize more
frequently used tasks by placing the corresponding tools at the top of
the tool list for the LLM \cite{shi2025judgingjudgessystematicstudy}.

\subsection{Providing Semantics for Super
Assistants}\label{providing-semantics-for-super-assistants}

The architecture described in this section provides a generic way to
expose the semantics of a GUI-based application in the form of tools and
to handle tool calls when they are sent from the assistant. A generic
OS-wide assistant can then execute a tool when a user\textquotesingle s
spoken request matches a function of the application. The generic
assistant can also use this information to call a sequence of functions
in an application to handle a user request that takes multiple steps.
Moreover, such a sequence may span multiple applications. For example,
if the user requests, ``pay my invoices for the last month,'' the
assistant could access the user's email application to retrieve invoices
from the past month, then proceed to the banking application to process
the payments.

Calls that span multiple applications require not only the execution of
steps but also the copying of information from one application to the
next. The tool result contains the information that the user sees on the
screen, in response to a tool call. For information-rich applications,
this should consist of a string representation of the entire screen
contents. If an application fails to provide enough information in
response to a tool call, the OS assistant may still attempt to parse the
relevant information from the GUI itself visually as a fallback.

A generic assistant can visually parse the GUI screen to extract its
semantics and plan the steps required to execute a
user\textquotesingle s spoken request. In theory, this allows it to
perform any task in any GUI-driven application without requiring that
application to be modified. However, semantic extraction is inherently
unreliable, as is the task performance ability of such generic
mechanisms.

Semantics provided by the application itself can be more precise and
reliable than those extracted from external visual observations by the
assistant. Therefore, an app that exposes high-quality semantics tends
to make a generic assistant have better task completion
\cite{lu2025axisefficienthumanagentcomputerinteraction, zhang2025apiagentsvsgui}.
This reliability is especially critical for speech interaction, where
ambiguity or misinterpretation can easily break the flow of the
conversation and undermine user trust.

\section{Local Model Evaluation}\label{local-model-evaluation}

Concerns about confidentiality, privacy, and data security often make
organizations hesitant to use AI assistants or language models hosted by
external vendors \cite{vanta_ai_governance}. To
retain full control over sensitive information and comply with internal
policies, many organizations would prefer to deploy their own assistants
using open-weight LLMs \cite{huang2025positiononpremisesllmdeployment, gaige2025open, moktali2024private}.

To assess the practical suitability of current open-weight LLMs for
speech-enabled, multimodal user interfaces, an evaluation was performed
across a set of tasks in a representative app based on the architecture
described above.

As opposed to typical agentic uses of function calling, in the functions
for speech enabling a GUI, most parameters are optional. A primary
reason for this is that a GUI stands between a user and (API) calls to
an underlying system. The user typically fills out only a subset of the
parameters in each turn, eventually filling out all parameters step by
step, leaving some defaults unchanged. Another difference is that
functions for GUI operation often contain many enumerations, so they
deserve much focus in an evaluation.

The most critical factor for speech interaction with a GUI is how
accurately the LLM maps a user expression onto a tool call to the GUI.
Inaccuracies require the user to correct the mistakes, which leads to
frustration. Too many mistakes make users reject the speech interface.
The second most important factor for the practical application of
LLM-based multimodal UIs is the response latency. In an interactive
environment, it is usually not acceptable if the response to a speech
request takes much longer than 3 seconds
\cite{fantinuoli2022definingmaximumacceptablelatency}. Some models are very accurate, but
slow and require significant compute.

Response time can be significantly decreased by employing more powerful
and expensive hardware. The required compute also increases when an
application has many users simultaneously. The expenses on hardware may
yield not choosing an open-weight LLM with the highest accuracy, but an
LLM that strikes a good balance between accuracy and compute.

\subsection{Method}\label{method}

A range of state-of-the-art LLMs was assessed, including both
proprietary models and small open-weight models (72B or smaller) with
varying parameter counts and training strategies (see Table
\ref{fig:modelEval}). The method outlined here is not meant as a
complete framework for function-calling evaluation. It serves merely to
measure the fit of an LLM for speech-enabling a specific GUI
application.

As a metric, the Berkeley Function-Calling Leaderboard (BFCL) V3 AST
evaluation was first considered
\cite{patil2025bfcl, rabinovich2025robustnessagenticfunctioncalling, Sreenivasan2024Unpacking}. BFCL's
strength is that it rigorously measures how accurately a language model
produces function calls that both adhere to a given JSON schema and
match ground-truth parameter values from the input. However, the test
set of BFCL only contains examples where \emph{all} parameters of a
supplied function are used. It lacks examples where certain parameters
should remain empty or be omitted. The strong contenders on the
leaderboard seem finetuned with that focus.

For the multimodal interaction as described in this article, screens or
panels function as user-facing proxies for making system calls.
Typically, most or all parameters of the functions exposed to the LLM
are optional, and not all parameters have a value in the ideal response.
When users voice a request, they usually provide only a subset of the
parameters for a screen or panel and augment them in subsequent turns,
either by voice or manually.

Another reason to choose a different metric than BFCL is that it
completely rejects model responses that deviate from the ideal response
by only one parameter value. Screens or panels typically contain a fair
number of parameters, and so do their corresponding functions. An error
in only one parameter is considered less severe than an error in all
parameters. Moreover, some user expressions are slightly ambiguous. It
was considered too much of a penalty to entirely reject a response
because a model does not fully capture the subtlety of the input
expression. To reflect this, the metric described in this section
averages the individual accuracies of each model response instead of
calculating the percentage of exactly correct responses.

Each model was evaluated using a set of 55 transcribed user requests
that mirrored core interaction patterns in the target multimodal user
interface. These were tested against 6 functions each, of which the main
ones had 5 or 6 parameters. To avoid making the task too easy or skewing
results positively, a highly diverse, free-form language set of examples
was used. All examples were in English and Dutch, where the English set
was a direct translation of the Dutch set. The tools used for the Dutch
examples were in Dutch, while the tools used for the English examples
were automatically translated into English and corrected for mistakes.
In principle, tools in English can be used even when the examples are in
another language. However, especially smaller models tend to perform
better when the function names, description, and parameters are in the
target language \cite{liu2025translationneedstudysolving, mondshine2025englishimpactprompttranslation}.

Some examples included self-corrections to test the models' ability to
handle revised user input. Others contained typical speech recognition
errors to test the models' ability to overcome those. The system prompt
read as follows:

\begin{codebox}\begin{minipage}{\columnwidth}\scriptsize\ttfamily
\begin{Verbatim}[breaklines=true, breaksymbol={}, breakindent=10pt]
You are a competent translator of user expression into tool calls for an app that records incidents in a data center.
- Record incident data strictly based on parameters provided by the user.
- Omit parameters from the response that the user has not provided, and NEVER make them up. If the information is not in the user message, it should NOT be in the response.
- Always return with tool calls and never ask for clarification or supplementary data.
- Be keen on self-correction within or between user messages. A parameter correction usually refers to the parameter uttered just before the correction itself.
\end{Verbatim}
\end{minipage}
\end{codebox}\strut
For each user request in the set, an ideal response was created as a
baseline. For each LLM, for each user request, the tool response
generated by the LLM was compared to the ideal response, and the
percentage of tool parameters, including the function name, that did not
match the ideal parameters was calculated. The average accuracy over the
entire set of 55 exemplary user requests was reported as the accuracy
for the tested LLM.

\begin{figure}
\centering
\includegraphics[width=\linewidth,height=3.55376in,keepaspectratio]{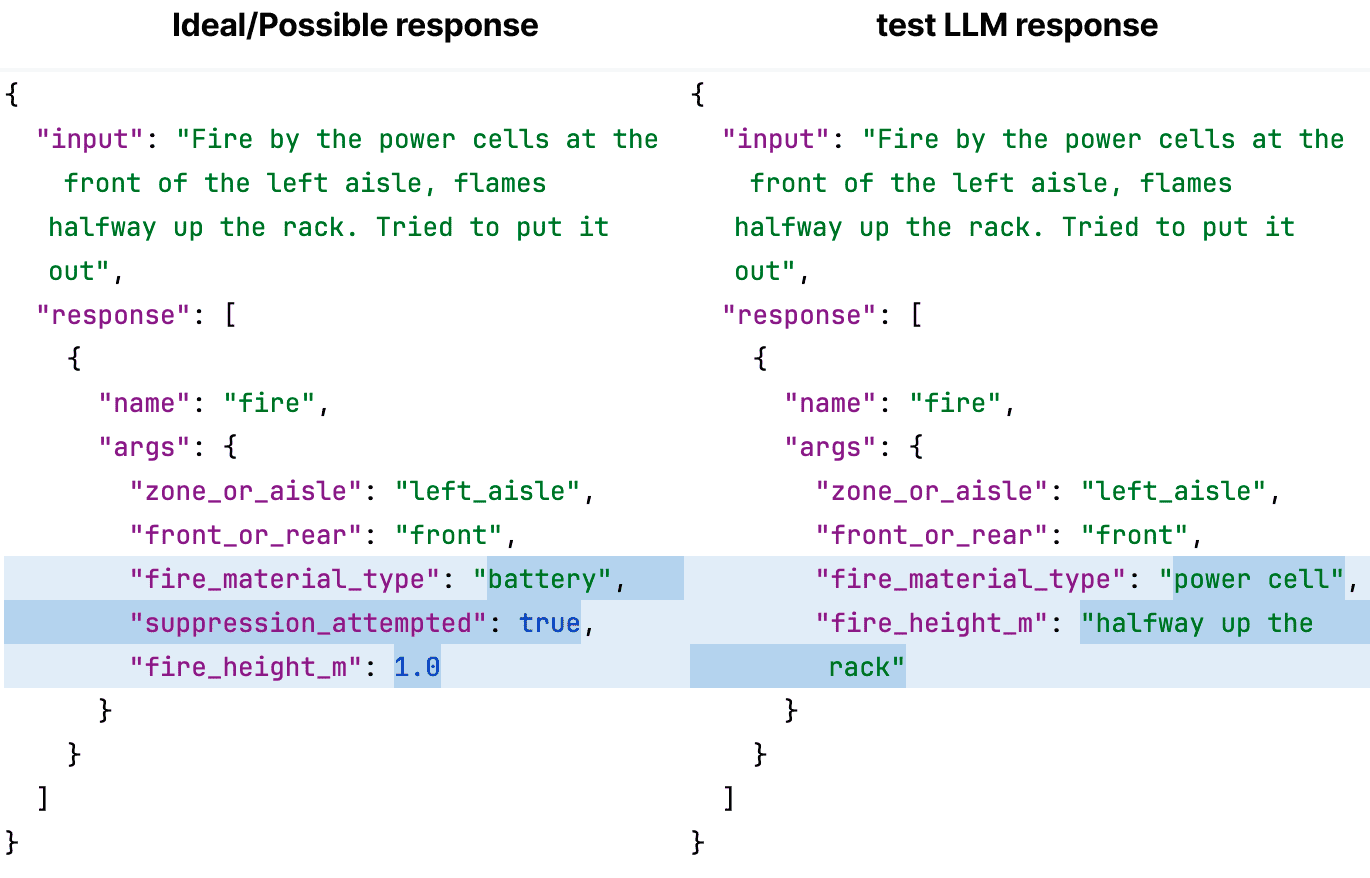}
\caption{Responses to a user expression in an incident app for a data
center facility. The ideal response is shown on the left; a faulty LLM
response is shown on the right. For this example, the tested model would
score 50\% accuracy in the current test metric. The total score of the
model is the average over all examples.}
\label{fig:responseComparison}
\end{figure}

To illustrate this, Figure \ref{fig:responseComparison} shows an
example comparison for a fire, reported in an incident management app
for a datacenter. The tested model on the right was rated a 50\%
accuracy for this example. The total number of values to compare was 6:
the function name plus 5 parameters. Of those, the tested LLM made 3
mistakes, so the accuracy for this expression on this LLM is 3/6 = 50\%.
A parameter like \emph{fire\_heigh\_m} is described in the tool
specification as the height of the fire in meters with \emph{type =
number}. In this case, the model is expected to guess a good numerical
value from ``halfway up the rack''. In the evaluation used here, the
``ideal'' response showed 1.0, which is roughly half the height of a
rack in a data center. If a model would output 0.5 or 1.5, this would
still be counted as correct, because this interpretation is subjective.
An output containing a literal textual value, however, is penalized
because it is not numerical, so it violates the schema. Similarly, in
the schema, the enumeration for \emph{fire\_material\_type} contained
``battery'' and not ``power cell'', so the output ``power cell'' is
considered an error, even though semantically it is correct. This method
allows for a practical assessment of each model's real-world
effectiveness within the target architecture.

Besides accuracy, for GUI interaction, the response time or latency of
models is also essential. The latency of a model is quite dependent on
the hardware on which the LLM is run. All open-weight models were run on
the same MacBook M3 with 128GB universal memory to make the latency
comparison fair. For strong contenders, the set was also run on an
NVIDIA H100 on RunPod to compare latencies.

The approach used in the current evaluation, while not based on a
standardized LLM benchmark, yields an indicative comparison tailored to
the demands of multimodal, speech-enabled user interfaces.

\subsection{Results}\label{results}

Table~\ref{fig:modelEval} presents the performance scores of the
evaluated models, reflecting their accuracy in translating user
expressions into tool calls and the average time for the model to
respond. For proprietary cloud models, the latency is not reported since
it is quite variable and less relevant for the current comparison. These
scores are displayed for both example sets, the Dutch and the English.
In general, there were minor differences between Dutch and English
scores. The original expectation was that for the open weight models,
the English experiment version would strongly outperform the Dutch,
because all models have been trained more extensively on English
language interpretation. Multiple runs have been tried, with slight
variations to the system prompt, function and parameter descriptions and
names. This showed that smaller models are very sensitive to such
variations. From extensive manual comparison of the errors in English
test cases to the same cases in Dutch, it seemed that the English cases
were slightly more prone to hallucination. However, the data was too
sparse to draw a conclusion on this subject. A more extensive evaluation
of multilingual performance can, e.g., be found in
\cite{islam2025llmshallucinatelanguagesmultilingual}. The average latency for the Dutch
example runs was usually slightly higher than for the English.

\begin{table*}[htbp]
    \centering
    \caption{LLM performance evaluation on translating user expressions to GUI tool calls. Average latency was measured for open-weight models.}
    \label{fig:modelEval}
    \renewcommand{\arraystretch}{1.2}
    \begin{tabular}{@{}llrrrrr@{}}
    \toprule
    \textbf{Model} & \textbf{Hardware} & \textbf{Size (GB)} & \multicolumn{2}{c}{\textbf{Accuracy (\%)}} & \multicolumn{2}{c}{\textbf{Latency (s)}} \\
    \cmidrule(lr){4-5}
    \cmidrule(lr){6-7}
    & & & \textbf{English} & \textbf{Dutch} & \textbf{English} & \textbf{Dutch} \\
    \midrule
    GPT 4.1 & - & - & 97.2 & 96.5 & - & - \\
    Claude Sonnet 4 & - & - & 97.0 & 97.6 & - & - \\
    GPT 4o & - & - & 96.3 & 96.8 & - & - \\
    GPT 5 & - & - & 95.8 & 96.4 & - & - \\
    Gemini 2.5 Pro & - & - & 95.3 & 94.4 & - & - \\
    Qwen3 32B Q8\_0 & M3 & 35 & 93.6 & 91.5 & 37.0 & 80.8 \\
    Qwen3 8B Q8\_0 & M3 & 9 & 91.7 & 91.5 & 20.2 & 15.7 \\
    Qwen3-32B & H100 & 65 & 89.6 & 92.3 & 12.6 & 14.7 \\
    OpenAI gpt-oss-20b & H100 & 14 & 89.2 & 91.4 & 1.5 & 1.7 \\
    Qwen3-8B-FP8 & H100 & 9 & 88.8 & 92.8 & 5.2 & 5.3 \\
    OpenAI gpt-oss-120b & H100 & 80 & 88.4 & 89.0 & 1.6 & 1.8 \\
    Llama 3.3 70B Instruct Q8\_0 & M3 & 74 & 86.2 & 89.7 & 9.0 & 12.6 \\
    Qwen3 1.7B Q8\_0 & M3 & 2.2 & 81.3 & 50.4 & 6.9 & 7.7 \\
    Qwen2.5 7B Instruct FP16 & M3 & 15 & 76.4 & 62.0 & 2.1 & 3.2 \\
    Ollama/OpenAI gpt-oss-120b & M3 & 65 & 74.8 & 74.6 & 5.2 & 6.1 \\
    Ollama/OpenAI gpt-oss-20b & M3 & 14 & 69.4 & 61.9 & 4.6 & 6.1 \\
    Qwen3 30B A3B Instruct FP16 & M3 & 61 & 69.1 & 63.7 & 1.7 & 2.0 \\
    xLAM 2 32B FC R FP16 & M3 & 65 & 48.9 & 75.2 & 19.2 & 23.0 \\
    Llama4 17B Scout 16E Instruct Q8\_0 & M3 & 116 & 44.0 & 58.6 & 4.8 & 7.0 \\
    \bottomrule
    \end{tabular}
    \end{table*}The results in Table \ref{fig:modelEval} indicate substantial variance
in task performance, with OpenAI's GPT-4.1 achieving the highest
accuracy, followed by other proprietary cloud models. The unquantized
Qwen3 32B showed a performance of nearly 94\% correctness in English,
which is the best accuracy score of the open-weight LLMs in the
experiment. More interesting for GUI interaction are gpt-oss-20/120b,
Llama 3.3 70B, and Qwen3 8B because they still have a fair accuracy but
are much faster than Qwen3 32B.

Hybrid Qwen3 models are slow and consume a lot of compute, even the
small 8B variant. They contain internal Chain of Thought (CoT) reasoning
steps before arriving at the answer \cite{yang2025qwen3technicalreport}.
When using tool calling, they tend to use these reasoning steps,
resulting in good accuracy. As a consequence, they have long response
times, which may require many steps and have variable and unpredictable
latencies. On a MacBook M3 with universal memory, the latencies for
Qwen3 thinking models are impractical for real-time GUI interaction.

Qwen3 32B FP16 was also tested on an H100 GPU on runpod, resulting in a
latency between 10 and 15 seconds, including the network call. Qwen3 8B
FP8 had a latency on H100 of around 5 seconds, but for some examples,
which were left out of the latency calculation, it got stuck in thinking
loops until the context window ran out. Such thinking loops produce
extensive latencies with empty responses at the end. Despite the
impressive performance in terms of accuracy of Qwen thinking models,
their latency may be impractical for real-time interactive use in GUIs.
The Qwen3 A3B instruct model, without reasoning, lacks the accuracy of
the ``thinking'' variants.

The OpenAI gpt-oss-20b and gpt-oss-120b performed well on an H100 on
RunPod\footnote{At the time of writing the gpt-oss models' tool calling
  had to be run through the OpenAI responses API on vLLM, because vLLM
  was not yet aligned with tool calling in the chat API using the OpenAI
  Harmony response format. VLLM is also not aligned yet to run these
  models on GPUs with lesser architectures than the H100.}, with a good
response time, including the network latency. Gpt-oss-20b outperforms
gpt-oss-120b: it has a higher accuracy, faster response, and a smaller
memory footprint. Given the combination of high accuracy and low
latency, gpt-oss-20b currently strikes the best balance for the type of
scenario described in this article.\\
gpt-oss-20b and gpt-oss-120b performed poorly on accuracy when run on a
Mac M3 with Ollama or LMStudio. The model is quantized using MXFP4. A
different type of quantization may have crippled the variants running on
Ollama on the Mac M3, and the model can hardly be coerced into strictly
following enumerations.

As shown in Table \ref{fig:modelEval}, the model xLAM-2-32b-fc-r, the
number 2 on the BFCL V3, did not perform well on the metric used in this
article, even though special attention was given to its system prompt
and it was run in FP16. It tended to hallucinate parameter values for
optional parameters that were present in the function schema, but that
were not inferable from the user expression at all. For booleans, most
models understand from the system prompt that boolean properties should
not be inferred to be \emph{false} if the user said nothing about that
property. Such parameters should be omitted from the response in that
case. xLAM-2-32b-fc-r tends to comply slightly better with the extra
omission encouragement mentioned for each boolean parameter. The
hallucination of the model would cause the GUI to incorrectly populate
parameters, leading to user frustration.

\subsection{Improving accuracy}\label{improving-accuracy}

The overall accuracy percentages in Table \ref{fig:modelEval} do not
always reflect the severity of a model\textquotesingle s inaccuracy.
Some inaccuracies are repairable using postprocessing, for instance:

\begin{itemize}
\item
  Omission of quotes around numbers.
\item
  Values of parameters set to ``null''.
\item
  Enumeration values are not followed strictly.
\end{itemize}

One systematic method for postprocessing faulty structured output by
LLMs is called Schema-Aligned Parsing (SAP)
\cite{gupta2024schema}, employed in
the BAML framework \cite{baml}. BAML applies a
parser that ``generously reads the output text and applies error
correction techniques with knowledge of the original schema''
\cite{gupta2024schema}.

In the experimental results, a mild type of inaccuracy occurred fairly
often, albeit not for Qwen3 32B: enumeration values are not strictly
followed, but are filled out using the exact wording of the user's
expression. An example of that is already shown in Figure
\ref{fig:responseComparison} where the user says `power cell' while the
enumeration contained ``battery''. The LLM filled out ``power cell'' for
the \emph{fire\_material\_type} parameter. In practice, such
inaccuracies can easily be detected because the parameter value deviates
from the values in the enumeration list. They can often be
programmatically corrected by applying synonym lists or other heuristic
methods. In this case, the synonym list would contain mappings where
``power cell'', ``power pack'', etc., can all be mapped to ``battery''.
This can be further improved by also including plurals of synonyms using
packages like spacy, inflect, or pluralize. One can even go one step
further by calculating the Levenshtein distance of the LLM output to
these alternatives to catch minor deviations, for instance, the
inclusion of articles like ``a'' or ``the''.

The system prompt has a great influence on the specifics of the tool
calling accuracy. This influence is quite model-specific, so it is
difficult to compare the influence of system prompt tuning
systematically. For an individual model, however, it is easy to run a
test set of utterances against a range of system prompts and iteratively
improve the prompt to coerce the model towards more desirable
performance \cite{kumar2025sculpt, zhang2024sprigimprovinglargelanguage}.

Higher accuracy in smaller models may be achieved by splitting the tool
calling process into two phases. The first phase only determines which
tool to call, and the second phase determines the parameter values of
that tool. This reduces the determination tasks, which tends to lead to
greater accuracy, especially in smaller models
\cite{mok2024llmbasedframeworksapiargument, varangotreille2025doinglesssurveyrouting, 10992798}

Higher accuracy can also be achieved by fine-tuning an open-weight model
on the specific functions and user expressions used in an application.
Fine-tuning for function calling is difficult and time-consuming. It
requires a study into a model's precise format, it requires an extensive
training, test, and evaluation set, and takes a lot of trial and error.
The rapid release sequence of improved open weight models may well
supersede accuracy improvements achieved by fine-tuning a model to a
specific application. However, in the later stages of
application-specific assistant development, it is well worth investing
in fine-tuning. The fine-tuning setup can then, with some modifications,
be reapplied to new, improved open-weight models that are released over
time.

\subsection{Summary}\label{summary}

The evaluation indicates that deploying open-weight LLMs for
speech-enabled, multimodal interfaces is feasible, provided suitable
hardware is used. Qwen3 32B offers high accuracy, but it is too slow for
practical interactive use, even on enterprise hardware. Models like
gpt-oss-20b and Llama 3.3 70B offer a more practical balance between
accuracy and latency, although even these models require
enterprise-grade GPUs for acceptable real-time responses. Many of the
observed inaccuracies, particularly in parameter formatting or
enumeration values, can be effectively addressed through post-processing
techniques such as schema-aligned parsing. Given the rapid pace of
progress in open-weight LLMs, the frequency of such glitches and the
resulting need for repair are expected to diminish further in the near
future.

\section{Conclusion}\label{conclusion}

The future of application development is just as much about tailoring a
GUI to the user as tailoring and exposing an app\textquotesingle s
capabilities via an API for OS-wide super assistants. Given the
widespread adoption of MCP among major players such as OpenAI, Google,
Microsoft, and X, MCP or a similar format is likely to become a
preferred way to expose an application\textquotesingle s capabilities to
a generic MLLM-based OS assistant. It is now time to start preparing the
navigational structures of apps to support the widespread take-off of
MLLM-based multimodal interaction.

The architecture introduced in this article is designed specifically to
support seamless speech-enabled interactions in GUIs. Central to this
design is the ViewModel, which is responsible for orchestrating context,
focus, and the exposure of functionality, ensuring comprehensive and
synchronized multimodal interaction. Crucially, this central role of the
ViewModel enables consistent multimodal feedback to enhance
conversational grounding and user understanding. Effective feedback
ensures that users know exactly how their spoken requests have been
interpreted and acted upon by the system. This is essential for building
trust and reliability in speech-driven interactions. The architecture
establishes a foundational layer that universally enables speech
coverage for basic functionalities, while also supporting sophisticated,
context-specific interactions. Explicitly exposing GUI semantics through
well-defined tools allows OS-level assistants to accurately interpret
and respond to complex user requests, even when such requests span
multiple applications. Providing semantics directly from the application
ensures greater precision and reliability and improves task completion
by generic assistants.

Evaluations indicate that open-weight local LLMs can handle many core
GUI tasks, but they remain less accurate than leading proprietary
models. Organizations that prioritize privacy and control may need to
accept some temporary reduction in performance if they choose local
deployment. Moreover, local deployment currently requires the use of
enterprise-grade hardware to achieve response times that are acceptable
in real-time interaction.

Achieving reliable multimodal interaction requires careful architectural
decisions that strike a balance between flexibility, maintainability,
and responsiveness. Developers and UX designers are encouraged to
proactively design new applications with built-in support for visual,
linguistic, and gesture-based interaction from the outset. By embracing
multimodal integration now, both existing and new software can fully
leverage the potential of advanced AI assistants, ensuring that future
systems are not only intuitive and versatile but also optimally prepared
for the next generation of super assistance.

\printbibliography
\typeout{get arXiv to do 4 passes: Label(s) may have changed. Rerun}
\end{document}